\documentclass[10pt,twocolumn,letterpaper]{article}
\usepackage{makecell}
\usepackage[final]{cvpr}
\usepackage{times}
\usepackage{epsfig}
\usepackage{graphicx}
\usepackage{amsmath,amssymb}
\usepackage{xcolor}
\usepackage{float}

\usepackage{booktabs}
\usepackage{multirow}
\usepackage{array}
\usepackage{url}
\usepackage{microtype}
\providecommand{\doi}[1]{\url{https://doi.org/#1}}
\usepackage{balance}
\usepackage{tikz}
\usetikzlibrary{arrows.meta}
\usepackage{pgfplots}
\pgfplotsset{compat=1.18}
\usepgfplotslibrary{groupplots}
\usetikzlibrary{positioning}

\usepackage{tikz}
\usetikzlibrary{arrows.meta, positioning, calc}
\usepackage[hidelinks]{hyperref}
\usepackage[nameinlink]{cleveref}
\usetikzlibrary{shapes.geometric,shapes.misc,arrows.meta}

\usepackage{tikz}
\usepackage{fontawesome5}
\usetikzlibrary{shapes.geometric, arrows.meta, positioning, calc, backgrounds, fit, shadows}

\usepackage{caption}

\usepackage{tikz}
\usetikzlibrary{
    positioning,
    calc,
    fit,
    arrows.meta,
    shapes.geometric
}

\begin{document}
\title{Identity by Design, Demographics by Accident: Demographic Leakage and Suppression in Behavioral Biometric Embeddings}

\author{
\centering
\parbox{0.9\textwidth}{\centering
Iyadh Khan$^{2}$,
Patrick Nilackshan$^{2}$,
Mohamed Aathif$^{2}$,
Mohamed Theesan$^{2}$\\[0.3em]
Sandareka Wickramanayake$^{2}$,
Sanka Rasnayaka$^{1}$,
Terence Sim Mong Cheng$^{1}$\\[0.8em]
$^{2}$Department of Computer Science and Engineering, University of Moratuwa, Sri Lanka\\
$^{1}$National University of Singapore, Singapore
}
}

\maketitle

\begin{abstract}

Behavioral biometric authentication (BBA) systems use deep learning models to transform biometric signals, such as eye movements, voice, keystroke/touchstroke dynamics, and gait, into identity embeddings for user authentication. While designed to encode identity, these embeddings may inadvertently reveal sensitive demographic attributes, including gender, age, and height. Consequently, an adversary with access to the authentication model can infer demographic information from biometric signals, including those of users unseen during training or enrollment. In this paper, we present the first systematic audit of demographic leakage in BBA systems, evaluating 11 models across 9 datasets spanning four biometric modalities. We further benchmark four post-hoc suppression methods-Incremental Variable Elimination (IVE), Hilbert–Schmidt Independence Criterion (HSIC), Adversarial Encoder-Decoder (AED), and Protected Attribute Suppression System (PASS) to assess their ability to mitigate demographic leakage while preserving authentication utility. Our analysis reveals substantial variation in leakage and suppressibility across modalities, model architectures, and learning objectives. While voice embeddings exhibit high leakage that can be effectively suppressed, keystroke/touchstroke embeddings exhibit lower leakage but are considerably more difficult to sanitize. These findings highlight a fundamental privacy risk in behavioral biometric authentication and provide insights into the factors governing demographic information suppressibility.

\end{abstract}
%--------------------------------------------------------------------
\section{Introduction}
%--------------------------------------------------------------------

Behavioral biometric authentication (BBA) systems authenticate users using signals such as eye gaze, voice, keystroke/touchstroke dynamics, and gait. Modern BBA systems transform raw biometric signals into compact identity embeddings that capture user-specific behavioral characteristics. Although optimized for identity discrimination, these embeddings may also encode demographic attributes unrelated to the authentication task\cite{kroger2020,rasnayaka2020}. As behavioral biometrics become increasingly deployed in applications such as healthcare, border control, and continuous authentication, understanding and mitigating such privacy risks is becoming increasingly important.

Previous studies have investigated demographic leakage from biometric signals and learned representations across modalities including eye gaze, voice, gait, and keystroke dynamics \cite{eberz2015,kroger2020,liu2019,makowski2020,abdrabou2024,zhang2018,bozkir2023,grandhi2025,steil2019,terhorst2019,rasnayaka2020}. However, these studies largely focuses on individual modalities, specific datasets, or particular privacy-preserving techniques, leaving the comparative behavior of demographic leakage and suppression in behavioral biometrics poorly understood.

To address this gap, we present the first systematic audit of demographic leakage and suppression in BBA systems. We consider a threat model in which an adversary extracts identity embeddings from a deployed authentication model and trains lightweight classifiers to infer sensitive attributes such as gender, age, and height. Because authentication models generalize across users, such inferences may also generalize to individuals not observed during training or enrollment. We evaluate 11 authentication models across 9 datasets spanning eye gaze, voice, keystroke/touchstroke, and gait, and benchmark four post-hoc suppression methods-IVE~\cite{terhorst2019}, HSIC~\cite{bortolato2020}, Adversarial Encoder-Decoder~\cite{noe2020}, and PASS~\cite{dhar2021} through their privacy-utility trade-offs.

\textbf{Contributions.} (1) We present the first systematic audit of demographic information leakage in behavioral biometric authentication systems across four biometric modalities, 11 authentication models, and 9 datasets. (2) We provide a unified analysis of the effects of modality, model architecture, and learning objectives on demographic leakage. (3) We benchmark four post-hoc suppression methods and evaluate their ability to reduce demographic leakage while preserving authentication utility. (4) We show that demographic suppressibility is governed primarily by the geometric organization of demographic information within the embedding space rather than by the magnitude of the initial leakage.

%--------------------------------------------------------------------
\section{Related Work}
%--------------------------------------------------------------------

\subsection{Privacy Leakage in Biometric Systems}
 Eye-gaze signals have been shown to expose
  a broad range of personal information~\cite{kroger2020,eberz2016,eberz2017,abdrabou2024},
  with privacy risks documented across authentication and VR
  contexts~\cite{bozkir2023,steil2019,grandhi2025,liu2019}. Gait
  encodes gender, age, height, and weight~\cite{rasnayaka2020}, while keystroke and behavioral dynamics
  carry user attributes that extend beyond authentication
  intent~\cite{senarath2023,nguyen2024,behaveformer2023}. In speaker
  verification and face recognition, embeddings trained purely for
  identity retain recoverable gender and age structure and ~\cite{noe2020,bortolato2020,dhar2021}
  Terh{\"o}rst et al.~\cite{terhorst2019} establish that face
  authentication embeddings retain demographic information even after the
  raw image is discarded, placing BBA embeddings too as the locus of leakage.

  Rasnayaka and Sim~\cite{rasnayaka2020} study privacy invasiveness of
  on-body IMU gait data, showing that gender, age, height, and weight are
  predictable from raw signals and hand-crafted features; they introduce
  the Privacy Vulnerability Index (PVI), and compare their gait results against published
  figures for face, fingerprint, and voice modalities. Their experiments
  nonetheless operate at the raw-signal level rather than on
  authentication embeddings, and propose no algorithmic suppression, only
  sensor-placement recommendations. Hanisch et al.~\cite{hanisch2025}
  survey anonymization techniques for behavioral biometrics by privacy
  goal and transformation type, but focus on raw-data anonymization
  without measuring embedding-level leakage. Melzi et al.~\cite{melzi2024}
  catalogue privacy-enhancing technologies for biometric authentication, cancelable biometrics etc. but evaluate them qualitatively without empirical leakage
  measurement or multi-modal comparison. None measures leakage directly
  from BBA embeddings, benchmarks suppression across
  heterogeneous behavioral biometrics, or evaluates the privacy-utility
  trade-off under a unified cross-modal protocol, the gap our work
  addresses.

\subsection{Demographic Suppression Methods}

Demographic suppression methods reduce attribute predictability from learned representations while retaining authentication utility \cite{dwgrl2023,dhar2021,terhorst2019,noe2020,bortolato2020}. Post-hoc methods operate on embeddings generated by the authentication model, avoiding retraining of the authentication pipeline.
Prior work evaluates suppression in isolated settings for one modality, one model family, or one method at a time, leaving open whether effectiveness generalizes across heterogeneous behavioral biometric embeddings, depends on authentication model design, and, crucially, whether suppressibility correlates with initial leakage magnitude or with the geometric organization of demographic information within the embedding space.

% Post-hoc methods transform  embeddings without retraining the authentication model; whereas in-training methods modify optimization and require training
% pipeline access. Prior work evaluates suppression in isolated settings for one modality, one model family, or one method at a time, leaving open questions like whether effectiveness generalizes across behavioral biometric embeddings or depends on authentication model design.

%--------------------------------------------------------------------
\section{Experimental Protocol}
%--------------------------------------------------------------------

\subsection{Experimental Overview}

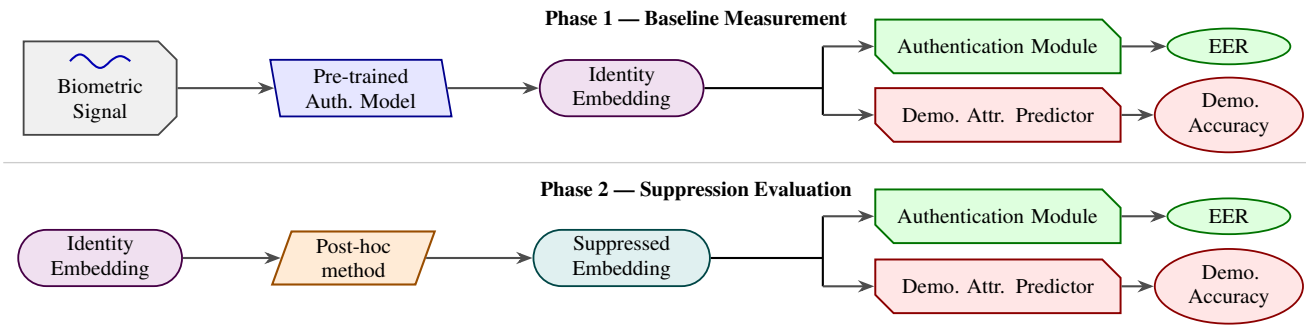
\begin{figure*}[t]
\centering
\resizebox{\textwidth}{!}{%
\begin{tikzpicture}[
  >=Stealth,
  every node/.style={font=\footnotesize},
  arr/.style={-{Stealth[length=2.2mm]}, thick, black!70},
  sbox/.style={draw=gray!55!black, line width=0.7pt, chamfered rectangle,
               chamfered rectangle corners={north east, south east},
               chamfered rectangle xsep=4pt, fill=gray!12, align=center,
               text width=1.66cm, minimum height=0.54cm, inner sep=1.6pt},
  mbox/.style={draw=blue!55!black, line width=0.7pt, trapezium,
               trapezium left angle=100, trapezium right angle=80,
               fill=blue!10, align=center, text width=1.82cm,
               minimum height=0.76cm, inner sep=2.4pt},
  ebox/.style={draw=violet!55!black, line width=0.7pt, rounded rectangle,
               fill=violet!12, align=center, text width=1.75cm,
               minimum height=0.68cm, inner sep=2.4pt},
  authevalbox/.style={draw=green!45!black, line width=0.75pt,
                     chamfered rectangle,
                     chamfered rectangle corners={north east, south west},
                     chamfered rectangle xsep=4pt, fill=green!13,
                     align=center, text width=1.72cm,
                     minimum height=0.72cm, inner sep=3pt},
  demoevalbox/.style={draw=red!55!black, line width=0.75pt,
                      chamfered rectangle,
                      chamfered rectangle corners={north east, south west},
                      chamfered rectangle xsep=4pt, fill=red!10,
                      align=center, text width=1.72cm,
                      minimum height=0.72cm, inner sep=3pt},
  phbox/.style={draw=orange!60!black, line width=0.8pt,
                trapezium, trapezium left angle=70,
                trapezium right angle=110, fill=orange!16,
                align=center, text width=1.48cm,
                minimum width=1.98cm, minimum height=0.70cm,
                inner sep=2.6pt},
  supbox/.style={draw=teal!55!black, line width=0.7pt, rounded rectangle,
                 fill=teal!12, align=center, text width=1.92cm,
                 minimum height=0.68cm, inner sep=2.4pt},
  eerbox/.style={draw=green!45!black, line width=0.7pt, ellipse,
                 fill=green!13, align=center, text width=1.04cm,
                 minimum height=0.48cm, inner sep=1.8pt},
  demobox/.style={draw=red!55!black, line width=0.7pt, ellipse,
                  fill=red!10, align=center, text width=1.28cm,
                  minimum height=0.54cm, inner sep=1.8pt},
]

%%──── PHASE 1 — Baseline Measurement ────────────────────────────────────
\node[font=\footnotesize\bfseries] at (-1.5, 1.14)
  {Phase 1 --- Baseline Measurement};

\node[sbox] (sig) at (-9.5, 0.2)
  {\tikz[baseline=-0.5ex,scale=0.45,
         every path/.style={blue!70!black,thick}]{%
     \draw (0,0) sin (0.3,0.2) cos (0.6,0) sin (0.9,-0.2) cos (1.2,0)
                 sin (1.5,0.15) cos (1.8,0);}\\[1pt]
   Biometric Signal};
\node[mbox] (mod)  at (-6.0, 0.2)
  {Pre-trained Auth.\ Model};
\node[ebox] (emb1) at (-2.5, 0.2)
  {Identity\\Embedding};

\draw[arr] (sig)  -- (mod);
\draw[arr] (mod)  -- (emb1);

\draw[thick] (emb1.east) -- (0.2, 0.2);
\draw[thick] (0.2, 0.76) -- (0.2, -0.16);

\node[authevalbox,text width=2.8cm] (authmod) at (2.55, 0.76)
  {Authentication Module};
\node[eerbox] (authout) at (5.65, 0.76)
  {EER};
\draw[arr] (0.2, 0.76) -- (authmod.west);
\draw[arr] (authmod.east) -- (authout.west);

\node[demoevalbox,text width=2.8cm] (demomod) at (2.55, -0.16)
  {Demo.\ Attr. Predictor};
\node[demobox] (demoout) at (5.65, -0.16)
  {Demo.\ Accuracy};
\draw[arr] (0.2, -0.16) -- (demomod.west);
\draw[arr] (demomod.east) -- (demoout.west);

%%──── HORIZONTAL DIVIDER ─────────────────────────────────────────────────
\draw[gray!40, line width=0.5pt] (-10.8, -0.8) -- (6.9, -0.8);
\node[font=\footnotesize\bfseries] at (-1.5, -1.18)
  {Phase 2 --- Suppression Evaluation};

%%──── POST-HOC ROW ───────────────────────────────────────────────────────
\node[ebox]   (emb2)   at (-9.5, -2.08)
  {Identity\\Embedding};
\node[phbox]  (phmeth) at (-6.1, -2.08)
  {Post-hoc\\method};
\node[supbox] (supemb) at (-2.5, -2.08)
  {Suppressed\\Embedding};

\draw[arr] (emb2)   -- (phmeth);
\draw[arr] (phmeth) -- (supemb);

\draw[thick] (supemb.east) -- (0.2, -2.08);
% \draw[thick] (0.2, -1.92)  -- (0.2, -3.06);
\draw[thick] (0.2, -1.52)  -- (0.2, -2.46);

\node[authevalbox,text width=2.8cm] (pauth) at (2.55, -1.52)
  {Authentication Module};
\node[eerbox] (pauthout) at (5.65, -1.52)
  {EER};
\draw[arr] (0.2, -1.52) -- (pauth.west);
\draw[arr] (pauth.east) -- (pauthout.west);

\node[demoevalbox,text width=2.8cm] (pdemo) at (2.55, -2.46)
  {Demo.\ Attr. Predictor};
\node[demobox] (pdemoout) at (5.65, -2.46)
  {Demo.\ Accuracy};
\draw[arr] (0.2, -2.46) -- (pdemo.west);
\draw[arr] (pdemo.east) -- (pdemoout.west);

\end{tikzpicture}%
}
\caption{Overview of the evaluation protocol.}
\label{fig:overview}
\end{figure*}

We evaluate demographic leakage and suppression in BBA using a two-stage protocol (\cref{fig:overview}). In the first stage, given a BBA framework we input the biometric signal of a user and obtain the identity embedding of that user and demographic leakage is measured by training probe classifiers to predict sensitive attributes from those embeddings. In the second stage, we apply demographic suppression methods to the embeddings and re-evaluate both demographic leakage and authentication utility. This allows us to measure whether demographic information is encoded in the learned embeddings and whether it can be reduced without substantially degrading authentication utility.

\subsection{Datasets and Authentication Models}

We use 9 datasets spanning the 4 modalities. Eye-gaze experiments use GazeBase \cite{gazebase2020}, GazeBaseVR \cite{gazebasevr2022}, and the VR Privacy dataset from NUS. Voice experiments use VoxCeleb1 \cite{nagrani2017} for suppression evaluation. Keystroke/touchstroke experiments use FETA \cite{feta2022}, KeyRecs \cite{keyrecs2023}, and IKDD \cite{ikdd2024}. Gait experiments use NUS-IGD \cite{nusFeatureGait020} and OU-ISIR \cite{ngo2014,ngo2015}. Dataset characteristics are summarized in ~\Cref{tab:datasets}, including the number of subjects, signal variables, demographic annotations, and gender distributions. Gender is evaluated across all modalities. Age, height, and weight are evaluated where annotations are available.

For each modality, we evaluate trained authentication models. Eye-gaze experiments use EyeKnowYouToo (EKYT) \cite{lohr2020,lohr2022a,lohr2022b,raju2024} and DeepEyedentification (DeepEye) \cite{makowski2021}, allowing comparison between metric-learning and classification-based embedding spaces. Voice experiments use ECAPA-TDNN \cite{desplanques2020} and WavLM \cite{chen2022wavlm} embeddings. Keystroke/touchstroke experiments use BehaveFormer \cite{behaveformer2023} and Type2Branch \cite{gonzalez2025}, including unimodal touch input and multimodal touch--IMU input settings where available. Gait experiments use recurrent, convolutional, and Transformer-based IMU authentication models. The purpose of this model set is not to optimize a single biometric system, but to compare demographic leakage across heterogeneous embedding spaces learned under different modalities, architectures, and training objectives.

\begin{table}[!t]
\centering
\caption{Dataset characteristics.
G=Gender, A=Age, H=Height, W=Weight.
Features: Acc.=Accelerometer,
Gyro.=Gyroscope, Time=Keystroke timing, IMU=Inertial Measurement Unit.\\
${}^{*}$ denotes a private dataset}
\label{tab:datasets}

{
\small
\setlength{\tabcolsep}{2pt}
\renewcommand{\arraystretch}{1.08}

\begin{tabular*}{\columnwidth}{@{\extracolsep{\fill}}llrllr@{}}
\toprule
\textbf{Dataset} & \textbf{Biometric} & \textbf{Subj.} & \textbf{Features} & \textbf{Demo.} & \textbf{M/F} \\
\midrule

GazeBase        & Gaze & 322 & Gaze(x,y)   & G       & 171/151 \\
GazeBaseVR     & Gaze & 407 & Gaze(x,y,z) & G       & 188/216 \\
VR Privacy$^{*}$& Gaze & 96  & Gaze,Pupil  & G       & 55/41 \\

\midrule

VoxCeleb1       & Voice & 1251 & Audio & G & 690/561 \\

\midrule

FETA            & Touch & 486 & Time,IMU & G,A & 231/185 \\
KeyRecs         & Touch & 100 & Time     & G,A & 60/39 \\
IKDD            & Touch & 164 & Time     & G,A & 88/76 \\

\midrule

NUS-IGD         & Gait & 53  & Acc.,Gyro.  & G,A  & 30/23 \\
OU-ISIR         & Gait & 640 & Acc.        & \shortstack[l]{H,W\\G,A}    & 320/320 \\
\bottomrule
\end{tabular*}
}
\end{table}

\subsection{Threat Model and Leakage Probing}

\begin{figure}[!t]
\centering
\resizebox{\columnwidth}{!}{%
\begin{tikzpicture}[
  font=\sffamily\small,
  every node/.style={align=center},
  >=Stealth,
  panel/.style={draw=gray!40, rounded corners=8pt, fill=gray!5, thick},
  modelshape/.style={draw=blue!55!black, line width=0.7pt, trapezium,
    trapezium left angle=100, trapezium right angle=80,
    fill=blue!10, align=center, text width=0.90cm,
    minimum height=0.62cm, inner sep=0.6pt},
  processshape/.style={draw=orange!65!black, very thick, circle, fill=orange!16,
    minimum size=1.7cm, inner sep=1pt, text width=1.35cm},
  predictorshape/.style={draw=red!55!black, line width=0.75pt,
    chamfered rectangle,
    chamfered rectangle corners={north east, south west},
    chamfered rectangle xsep=4pt, fill=red!10,
    align=center, text width=1.28cm,
    minimum height=0.62cm, inner sep=2.2pt},
  arr/.style={-{Stealth[length=2.0mm, width=1.7mm]}, thick, black!80},
  cap/.style={font=\sffamily\tiny\itshape, text=black!60},
  stagetitle/.style={font=\sffamily\bfseries\small, anchor=north west, text=black!80},
  iconlab/.style={font=\sffamily\footnotesize, text=black!70},
]

%% ============ STAGE 1 ============
\node[stagetitle] (s1title) at (0,0) {Stage 1: Predictor Training};

% --- main row, all at y = -2.1 ---
\node (uicon1) at (0.55,-1.20) {\Large\textcolor{black!75}{\faUsers}};
\node (sigicon1) at (0.55,-1.65) {\large\textcolor{teal!55!black}{\faWaveSquare}};
\node[iconlab] (siglabel1) at (0.55,-2.18) {Biometric\\signal};
\node[iconlab] at (0.85,-0.82) {Training Cohort};

\node[modelshape] (auth1) at (2.65,-1.65) {\scriptsize\textbf{Auth.\\[-2pt]Model}};

\node (embicon1) at (4.85,-1.65) {%
  \begin{tikzpicture}[baseline]
    \foreach \i/\c in {0/teal!35,1/teal!50,2/teal!65,3/teal!80} {
      \pgfmathsetmacro{\y}{-\i*0.13}
      \draw[draw=teal!60!black, fill=\c, rounded corners=1pt] (-0.30,\y) rectangle (0.30,\y+0.09);
    }
  \end{tikzpicture}};
\node[iconlab] at (4.85,-1.20) {Embeddings};

\node[iconlab] at (6.6,-1.85) {\textbf{Train Predictor}};
% \node[processshape] (train) at (6.55,-1.65) {\textbf{Train\\Predictor}\\[1pt]
%   {\footnotesize\textcolor{orange!65!black}{\faCogs}}};

\node[predictorshape] (clf1) at (8.85,-1.65) {\scriptsize\textbf{Predictor}};

% --- GT tag, placed safely between title and row, above train only ---
\node (gticon) at (6.55,-0.78) {\large\textcolor{black!65}{\faTag}};
\node[iconlab] at (6.55,-0.42) {Attribute Labels};

% --- arrows ---
\draw[arr] (sigicon1.east) -- (auth1.west);
\draw[arr] (auth1.east) -- (embicon1.west);
\draw[arr] (embicon1.east) -- (clf1.west);

\draw[thick, black!80] (gticon.south) -- (6.55,-1.65);

\node[inner sep=0pt, minimum width=1.45cm, minimum height=0pt] (s1right) at (9.5,-1.65) {};

\begin{scope}[on background layer]
\node[panel, fit=(s1title)(uicon1)(siglabel1)(gticon)(auth1)(embicon1)(clf1)(s1right), inner xsep=5pt, inner ysep=2pt] (panel1) {};
\end{scope}

%% ============ STAGE 2 ============
\begin{scope}[yshift=-2.95cm]
\node[stagetitle] (s2title) at (0,0) {Stage 2: Attribute Inference};

\node (uicon2) at (0.55,-1.20) {\Large\textcolor{black!75}{\faUser}};
\node (sigicon2) at (0.55,-1.75) {\large\textcolor{teal!55!black}{\faWaveSquare}};
\node[iconlab] (siglabel2) at (0.55,-2.28) {Biometric\\signal};
\node[iconlab] at (0.85,-0.72) {New User};

\node[modelshape] (auth2) at (2.65,-1.75) {\scriptsize\textbf{Auth.\\[-2pt]Model}};

\node (embicon2) at (4.85,-1.75) {%
  \begin{tikzpicture}[baseline]
    \foreach \i/\c in {0/teal!35,1/teal!50,2/teal!65,3/teal!80} {
      \pgfmathsetmacro{\y}{-\i*0.13}
      \draw[draw=teal!60!black, fill=\c, rounded corners=1pt] (-0.30,\y) rectangle (0.30,\y+0.09);
    }
  \end{tikzpicture}};
\node[iconlab] at (4.85,-1.30) {Embedding};

\node[predictorshape] (clf2) at (7.35,-1.75) {\scriptsize\textbf{Predictor}};

\node (out2) at (9.5,-1.75) {\Large\textcolor{violet!55!black}{\faIdBadge}};
\node[iconlab] (predlabel2) at (9.5,-2.35) {Predicted\\Attribute};

\draw[arr] (sigicon2.east) -- (auth2.west);
\draw[arr] (auth2.east) -- (embicon2.west);
\draw[arr] (embicon2.east) -- (clf2.west);
\draw[arr] (clf2.east) -- (out2.west);

\begin{scope}[on background layer]
\node[panel, fit=(s2title)(uicon2)(siglabel2)(auth2)(embicon2)(clf2)(out2)(predlabel2), inner xsep=5pt, inner ysep=2pt] (panel2) {};
\end{scope}
\end{scope}

\end{tikzpicture}
}
\caption{Two-stage threat model for demographic attribute inference from a frozen authentication model.}\label{fig:threat_model}
\end{figure}
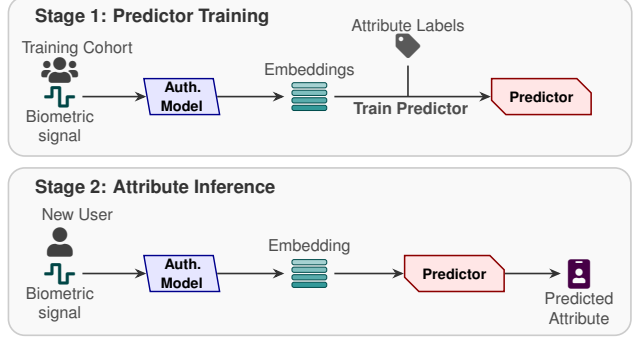

Let $M:\mathcal{X}\rightarrow\mathcal{R}$ denote a deployed BBA model that maps a biometric signal $X\in\mathcal{X}$ to an identity embedding $R=M(X)\in\mathcal{R}$. We consider a \emph{passive adversary} who has access to the authentication model $M$ and a labeled reference dataset $\mathcal{D}*{\mathrm{ref}}={(X_i,A_i)}*{i=1}^{N}$, where $X_i$ denotes a biometric sample and $A_i$ denotes a demographic attribute such as gender, age, or height. The adversary generates embeddings $R_i=M(X_i)$ and trains a demographic probe $f:\mathcal{R}\rightarrow\mathcal{A}$ to predict attribute values from the embedding space (\Cref{fig:threat_model}).

After training, the adversary applies the probe to embeddings extracted from previously unseen users. Since BBA models are designed to generalize across users, demographic inference may also generalize to individuals who were not observed during model training or enrollment. The reference dataset may be independently collected, obtained from a public dataset, or consist of embeddings generated using the same authentication pipeline.

To quantify demographic leakage, we train a set of standard probe classifiers, including Logistic Regression, Random Forest, XGBoost, and Multi-Layer Perceptron (MLP), and report the performance of the strongest probe as a best-effort estimate of demographic recoverability from the learned representation. For binary attributes, leakage is measured relative to the $50\%$ chance baseline, noting that below-chance predictions can be inverted. For multi-class attributes, including discretized age and height, leakage is assessed relative to the majority-class baseline.

\subsection{Suppression Methods}

We evaluate four post-hoc suppression methods which operate directly on embeddings. The post-hoc methods include Incremental Variable Elimination (IVE)~\cite{terhorst2019}, which removes attribute-associated principal directions; Hilbert--Schmidt Independence Criterion (HSIC)~\cite{bortolato2020}, which reduces statistical dependence between embeddings and sensitive attributes; Protected Attribute Suppression System (PASS)~\cite{dhar2021}, which suppresses demographic predictability while preserving embedding geometry; and Adversarial Encoder-Decoder (AED)~\cite{noe2020}, which uses adversarial reconstruction to remove attribute information while retaining identity-relevant structure.

 % Stage 1 trains an attribute predictor $P^{*}$ from embeddings and attribute labels; Stage 2 uses $P^{*}$ to infer attributes from a new user's embedding.

\subsection{Evaluation Metrics}

Authentication utility is measured using equal error rate (EER), where lower values indicate better authentication utility. Privacy leakage is measured using demographic probe accuracy before and after suppression. 

We also report the Privacy Vulnerability Index (PVI)~\cite{rasnayaka2020}. In its general form, PVI combines the predictability of each private attribute with its perceived importance:
\begin{equation}
\text{PVI} = \frac{\sum_i s_i p_i}{\sum_i s_i},
\end{equation}
where $p_i$ measures how well attribute $i$ can be inferred and $s_i$ denotes the importance weight assigned to that attribute. Since gender is the common attribute evaluated across all modalities in this work, we use an attribute-level PVI where $p_i$ is the gender probe recoverability:
\begin{equation}
\text{PVI}_{gender} =
\max(\text{gender accuracy}, 1-\text{gender accuracy}).
\end{equation}
Thus, 50\% indicates chance-level gender recoverability, while higher values indicate stronger privacy leakage.

A useful suppression method should reduce demographic accuracy and PVI toward chance while preserving authentication utility. To summarize the privacy-utility trade-off after suppression, we use the Privacy Gain and Identity Loss
Coefficient (PIC)~\cite{terhorst2019}. For gender, let $fic$ denote demographic
inference error, computed as $1-\text{gender accuracy}$. Given baseline values $(fic, eer)$ and post-suppression
values $(fic', eer')$, PIC is defined as
\begin{equation}
\text{PIC} = \frac{fic' - fic}{fic} - \frac{eer' - eer}{eer}.
\end{equation}

Positive PIC indicates that the relative privacy gain exceeds the relative
authentication cost, while negative PIC indicates an unfavorable trade-off. ~\Cref{fig:heatmap} summarizes PIC across the post-hoc suppression
configurations, and ~\Cref{fig:pvi_all_modalities} plots PVI against EER
to visualize the broader privacy--utility frontier.

%--------------------------------------------------------------------
\section{Results and Analysis}
\label{sec:results}
%--------------------------------------------------------------------

This section answers five research questions: (1) Is demographic leakage common across behavioral biometric modalities? (2) How much does the model training objective or loss function affect leakage severity? (3) Which suppression methods are effective, and how strongly does their effect depend on the underlying authentication model? (4) Does multimodal input encode more demographic structure and make suppression harder? (5) Can suppression techniques originally developed for other biometric domains generalize to behavioral biometric embeddings? We conclude with a privacy–utility evaluation using Privacy Gain and Identity Loss Coefficient.

\subsection{Demographic Leakage Across Behavioral Biometric Modalities}

\begin{table}[!t]
\centering
\caption{Baseline leakage and authentication utility. Chance gender recoverability is 50\%; “*" indicates Classification accuracy reported only by DeepEye instead of EER.}
\label{tab:baseline}

\small
\renewcommand{\arraystretch}{1}
\setlength{\tabcolsep}{4pt}

\resizebox{\columnwidth}{!}{%
\begin{tabular}{llllr}
\toprule
\textbf{Modality} & \textbf{Dataset} & \textbf{Model} & \textbf{EER/Acc*} & \textbf{Gen.}\\
\midrule
\multirow{6}{*}{Eye Gaze}
 & \multirow{2}{*}{GazeBase}
   & EKYT & 8.77 & 77.8 \\
 & & DeepEye & 90.22* & 57.6 \\
\cmidrule{2-5}
 & \multirow{2}{*}{GazeBaseVR}
   & EKYT & 7.34 & 65.5 \\
 & & DeepEye & 93.62* & 61.8 \\
\cmidrule{2-5}
 & \multirow{2}{*}{\shortstack[l]{VR Privacy}}
   & EKYT & 6.67 & 65.8 \\
 & & DeepEye & 93.88* & 56.9 \\
\midrule
\multirow{2}{*}{Voice}
 & \multirow{2}{*}{VoxCeleb1}
   & ECAPA-TDNN & 0.90 & 86.1 \\
 & & WavLM & 4.64 & 94.8 \\
\midrule
\multirow{6}{*}{Keystroke}
 & \multirow{2}{*}{FETA}
   & BehaveFormer & 15.42 & 61.8 \\
 & & Type2Branch & 8.07 & 61.3 \\
\cmidrule{2-5}
 & \multirow{2}{*}{KeyRecs}
   & BehaveFormer & 19.53 & 69.1 \\
 & & Type2Branch & 21.76 & 66.7 \\
\cmidrule{2-5}
 & \multirow{2}{*}{IKDD}
   & BehaveFormer & 24.74 & 66.7 \\
 & & Type2Branch & 22.0 & 67.1 \\
\midrule
\multirow{10}{*}{Gait}
 & \multirow{5}{*}{NUS-IGD}
   & whuGAIT & 24.27 & 50.2 \\
 & & FG-FCN & 0.62 & 70.4 \\
 & & FG-RN & 2.08 & 56.0 \\
 & & ST & 5.44 & 50.6 \\
 & & ACT & 4.69 & 67.1 \\
\cmidrule{2-5}
 & \multirow{5}{*}{OU-ISIR}
   & whuGAIT & 35.38 & 61.9 \\
 & & FG-FCN & 5.64 & 70.4 \\
 & & FG-RN & 9.39 & 69.5 \\
 & & ST & 16.58 & 63.7 \\
 & & ACT & 14.91 & 67.3 \\
\midrule
\end{tabular}%
}
\end{table}

~\Cref{tab:baseline} summarizes baseline authentication utility and demographic leakage before suppression. Across the evaluated modalities, authentication embeddings expose recoverable demographic information. Evidence of leakage appears across modalities: GazeBase EKYT reaches 77.8\% gender accuracy, OU-ISIR gait models reach up to 70.4\%, keystroke embeddings reach up to 69.1\%, and voice embeddings show the strongest leakage, with 86.09\% for ECAPA-TDNN and 94.76\% for WavLM. These results show that demographic leakage is not confined to a single behavioral modality; it appears across heterogeneous authentication embeddings, although its magnitude depends on the dataset and the type of authentication model used.

\subsection{Effect of Training Objective and Embedding Geometry}

We examine whether leakage severity changes with how the authentication embedding is learned. Eye Gaze provides a direct comparison between EKYT, which uses
a metric-learning-oriented representation, and DeepEyedentification, which uses
closed-set classification. Across all three gaze datasets, EKYT shows higher baseline gender leakage than DeepEyedentification: 77.8\% versus 57.6\% on GazeBase, 65.5\% versus 61.8\% on GazeBaseVR, and 65.8\% versus 56.9\% on VR Privacy. This consistent pattern suggests that metric-learning-oriented embeddings can expose stronger demographic separability when identity-discriminative structure aligns with demographic variation.

Gait results provide further evidence that model design changes demographic leakage within the same modality. On OU-ISIR, FG-FCN and FG-RN leak gender at 70.43\% and 69.50\%, respectively, while whuGAIT leaks at 61.93\%. On NUS-IGD, FG-FCN reaches 70.40\% gender accuracy, while whuGAIT and the standard Transformer remain near chance. Since these models operate on the same modality, the differences point to the role of architecture and representation learning in shaping demographic structure.
Overall, these results show that demographic leakage is not determined by modality alone. Training objective, architecture, dataset, and embedding geometry all influence how much demographic information is encoded in the authentication embeddings. In the gaze experiments, metric-learning-oriented embeddings expose stronger gender leakage than closed-set classification embeddings, while the gait experiments show that different architectures trained on the same modality can produce different leakage levels. These geometry differences foreshadow a key finding explored below: models whose
embedding geometry concentrates demographic information in a separable subspace will
also be more amenable to suppression, independently of their absolute leakage level.

\begin{table}[!t]
\centering
\caption{Keystroke Demographic Suppression; “↑” indicates increased leakage}
\label{tab:keystroke_all}

\small
\setlength{\tabcolsep}{1.5pt}
\renewcommand{\arraystretch}{0.92}

\begin{tabular}{@{}
>{\raggedright\arraybackslash}p{0.35\columnwidth}
>{\raggedright\arraybackslash}p{0.20\columnwidth}
>{\raggedleft\arraybackslash}p{0.13\columnwidth}
>{\raggedleft\arraybackslash}p{0.13\columnwidth}
>{\raggedleft\arraybackslash}p{0.13\columnwidth}@{}}
\toprule
\textbf{Model} & \textbf{Method}
& \textbf{EER}
& \textbf{Gen.} & \textbf{Age} \\
\midrule

% ---------------- FETA ----------------
\multicolumn{5}{l}{\textbf{Dataset: FETA}} \\
\cmidrule(lr){1-5}
\multirow{5}{*}{\shortstack[l]{BehaveFormer\\(Touch)}}
& Baseline      & 15.42 & 61.97 & 58.82 \\
& PASS        & 17.15 & 57.84 & 55.88 \\
& AED         & 16.16 & 50.00 & 57.35 \\
& HSIC        & 24.23 & 50.00 & 58.82 \\
& IVE         & 35.00 & 55.39 & 55.88 \\
\cmidrule(lr){1-5}

\multirow{5}{*}{\shortstack[l]{BehaveFormer\\(Touch+IMU)}}
& Baseline      & 12.24 & 63.52 & 60.29 \\
& \textbf{PASS} & 12.25 & \textbf{56.86} & 55.88 \\
& AED         & 13.01 & 53.43 & 58.82 \\
& HSIC        & 34.95 & 51.96 & 51.47 \\
& IVE         & 44.71 & 55.39 & 44.12 \\
\cmidrule(lr){1-5}

\multirow{5}{*}{\shortstack[l]{Type2Branch\\(Touch)}}
& Baseline      & 8.07  & 62.68 & 64.22 \\
& PASS        & 10.34 & 60.29 & 59.31 \\
& AED         & 11.58 & 65.20$\uparrow$ & 57.35 \\
& HSIC        & 19.39 & 63.24$\uparrow$ & 51.47 \\
& IVE         & 26.56 & 59.31 & 46.57 \\
\cmidrule(lr){1-5}

\multirow{5}{*}{\shortstack[l]{Type2Branch\\(Touch+IMU)}}
& Baseline      & 6.86  & 65.49 & 66.18 \\
& PASS        & 10.11 & 58.33 & 63.24 \\
& \textbf{AED} & 10.98 & \textbf{59.31} & 52.94 \\
& HSIC        & 11.63 & 57.35 & 61.27 \\
& IVE         & 26.55 & 53.92 & 50.49 \\
\midrule

% ---------------- KeyRecs ----------------
\multicolumn{5}{l}{\textbf{Dataset: KeyRecs}} \\
\cmidrule(lr){1-5}
\multirow{6}{*}{\shortstack[l]{BehaveFormer}}
& Baseline         & 19.53 & 69.05 & 59.52 \\
& PASS           & 20.87 & 64.29 & 59.52 \\
& AED            & 20.34 & 61.90 & 64.29$\uparrow$ \\
& HSIC           & 20.06 & 64.29 & 57.14 \\
& \textbf{IVE}   & 35.47 & \textbf{50.00} & 59.52 \\
\cmidrule(lr){1-5}

\multirow{6}{*}{\shortstack[l]{Type2Branch}}
& Baseline         & 21.76 & 66.67 & 52.38 \\
& \textbf{PASS}  & 29.01 & \textbf{57.14} & 52.38 \\
& AED            & 22.92 & 64.29 & 50.00 \\
& HSIC           & 24.48 & 66.67 & 45.24 \\
& IVE            & 37.80 & 66.67 & 50.00 \\
\midrule

% ---------------- IKDD ----------------
\multicolumn{5}{l}{\textbf{Dataset: IKDD}} \\
\cmidrule(lr){1-5}
\multirow{6}{*}{\shortstack[l]{BehaveFormer}}
& Baseline         & 24.74 & 66.67 & 61.43 \\
& PASS           & 26.12 & 64.90 & 60.12 \\
& AED            & 28.01 & 61.43 & 55.00 \\
& HSIC           & 27.90 & 62.43 & 54.50 \\
& \textbf{IVE}   & 26.43 & \textbf{54.76} & 56.67 \\
\cmidrule(lr){1-5}

\multirow{6}{*}{\shortstack[l]{Type2Branch}}
& Baseline         & 22.00 & 67.14 & 61.43 \\
& \textbf{PASS}  & 24.85 & \textbf{63.81} & 59.76 \\
& AED            & 26.30 & 64.29 & 57.14 \\
& HSIC           & 25.77 & 65.52 & 55.62 \\
& IVE            & 26.43 & 59.28 & 54.29 \\
\bottomrule
\end{tabular}
\end{table}

\subsection{Suppression Effectiveness Depends on the Underlying Model}

We investigate whether suppression effectiveness depends on the authentication model. Rather than identifying a universally effective suppressor, we observe strong method-by-model interactions. Each suppression method reduces demographic leakage in some settings, but can provide limited benefit, increase leakage, or degrade authentication utility in others.

PASS is most effective when the embedding geometry aligns with its preservation objective. In GazeBase EKYT, it reduces gender leakage from 77.8\% to 69.9\% while preserving EER, making it the strongest post-hoc method in that setting. However, it is less reliable for several classification-based gaze and gait embeddings, suggesting that preserving embedding geometry is beneficial only when the original geometry already supports suppression; otherwise, it may also preserve demographic structure when identity and demographic information are entangled.

HSIC performs well for several classification-based and convolutional embeddings, reducing leakage close to chance in GazeBase DeepEyedentification and from 70.40\% to 57.96\% in NUS-IGD FG-FCN. However, these privacy gains can come at the cost of substantially higher EER when demographic and identity information are strongly coupled, particularly in keystroke settings.

IVE is effective when demographic information lies in removable linear directions. It performs well in selected gaze and keystroke settings but is often the most utility-sensitive method. For example, in FETA multimodal BehaveFormer, EER increases from 12.24\% to 44.71\%, indicating that removing demographic directions can also eliminate identity-discriminative information.

AED achieves some of the strongest leakage reductions, particularly for gait and voice embeddings. It lowers NUS-IGD FG-FCN gender leakage from 70.40\% to 61.97\% and reduces voice leakage close to chance with minimal impact on EER. However, its effectiveness remains model dependent, reflecting whether adversarial reconstruction can remove demographic information without disrupting identity-discriminative structure.

Overall, suppression effectiveness is determined not only by the suppression method or the initial leakage magnitude, but by the geometric organization of the embedding space. High-leakage voice embeddings (86-94\%) can be almost fully suppressed, whereas moderate-leakage keystroke and touchstroke embeddings (62-69\%) remain resistant across all four methods. These findings suggest that demographic suppression should not be treated as a universal preprocessing step. Instead, suppression strategies should be empirically validated for each embedding space, with particular attention to whether demographic information occupies a separable subspace or is inherently entangled with identity-discriminative structure.

\subsection{Effect of Multimodal Input}

\begin{table}[!t]
\centering
\caption{ Voice Demographic Suppression; “↑” indicates increased leakage}
\label{tab:voice}
\small
\setlength{\tabcolsep}{2pt}
\renewcommand{\arraystretch}{0.84}
\resizebox{\columnwidth}{!}{%
\begin{tabular*}{\columnwidth}{@{\extracolsep{\fill}}l l r r@{}}
\toprule
\textbf{Model} & \textbf{Method}
& \textbf{EER} & \textbf{Gen.} \\
\midrule
\multicolumn{4}{l}{\textbf{Dataset: VoxCeleb1 Dataset}} \\
\cmidrule(lr){1-4}
\multirow{5}{*}{ECAPA-TDNN}
& Baseline & 0.90 & 86.1 \\
& IVE & 1.04 & 75.6 \\
& HSIC & 2.66 & 54.1 \\
& PASS & 1.37 & 70.8 \\
& \textbf{AED} & \textbf{0.97} & \textbf{52.1} \\
\cmidrule(lr){1-4}
\multirow{5}{*}{WavLM}
& Baseline & 4.64 & 94.8 \\
& IVE & 4.81 & 77.6 \\
& HSIC & 4.74 & 82.4 \\
& PASS & 4.77 & 88.9 \\
& \textbf{AED} & \textbf{4.84} & \textbf{57.0} \\
\bottomrule
\end{tabular*}%
}
\end{table}

We hypothesized that adding an additional sensing modality would improve authentication utility but could also introduce more demographic information into the embedding space. The \Cref{tab:keystroke_all} FETA results support this hypothesis. Compared with touch-only embeddings, embeddings that incorporate inertial measurements improve authentication for both BehaveFormer and Type2Branch, with lower EER. However, the same multimodal embeddings also show higher gender and age leakage, indicating that the added sensor information contributes demographic structure as well as identity-discriminative information.

The suppression results further show that this added structure is harder to remove without affecting authentication. For BehaveFormer, PASS and AED reduce gender leakage while keeping EER changes relatively small compared with HSIC and IVE. In contrast, HSIC and IVE achieve stronger leakage reduction in some cases but substantially degrade authentication utility, with EER increasing to 34.95\% and 44.71\%, respectively. For Type2Branch, IVE gives the lowest gender leakage, but the associated EER increase makes the trade-off unfavorable.

These results show that multimodal inputs improve authentication but do not improve demographic privacy. Instead, inertial measurements can introduce additional demographic structure into the learned representation, making suppression more costly when that structure overlaps with identity-relevant information.

\subsection{Generalization of Suppression Methods to Behavioral Biometric Embeddings}

\begin{table}[!t]
\centering
\caption{Eye-gaze demographic suppression.“↑” indicates increased leakage. “*" indicates Classification accuracy reported only by DeepEye instead of EER.}
\label{tab:all_results}

\small
\setlength{\tabcolsep}{2pt}
\renewcommand{\arraystretch}{0.84}

\resizebox{\columnwidth}{!}{%
\begin{tabular*}{\columnwidth}{@{\extracolsep{\fill}}l l r r@{}}
\toprule
\textbf{Model} & \textbf{Method}
& \textbf{EER/Acc*} & \textbf{Gen.} \\
\midrule

\multicolumn{4}{l}{\textbf{Dataset: GazeBase}} \\
\cmidrule(lr){1-4}
\multirow{5}{*}{EKYT}
& Baseline & 8.77 & 77.8 \\
& IVE & 8.77 & 74.8 \\
& HSIC & 8.77 & 75.7 \\
& \textbf{PASS} & 8.77 & \textbf{69.9} \\
& AED & 9.10 & 71.5 \\
\cmidrule(lr){1-4}
\multirow{5}{*}{DeepEye}
& Baseline & 90.22* & 57.6 \\
& IVE & 79.78* & 54.3 \\
& \textbf{HSIC} & 86.96* & \textbf{51.9} \\
& PASS & 89.42* & 57.7$\uparrow$ \\
& AED & 79.02* & 56.8 \\
\midrule

\multicolumn{4}{l}{\textbf{Dataset: GazeBaseVR}} \\
\cmidrule(lr){1-4}
\multirow{5}{*}{EKYT}
& Baseline & 7.34 & 65.5 \\
& IVE & 5.42 & 55.0 \\
& HSIC & 6.16 & 59.5 \\
& PASS & 5.00 & 62.3 \\
& \textbf{AED} & 8.33 & \textbf{54.4} \\
\cmidrule(lr){1-4}
\multirow{5}{*}{DeepEye}
& Baseline & 93.62* & 61.8 \\
& \textbf{IVE} & 90.02* & \textbf{59.3} \\
& HSIC & 92.96* & 65.1$\uparrow$ \\
& PASS & 94.9* & 61.3 \\
& AED & 89.96* & 63.2$\uparrow$ \\
\midrule

\multicolumn{4}{l}{\textbf{Dataset: VR Privacy}} \\
\cmidrule(lr){1-4}
\multirow{5}{*}{EKYT}
& Baseline & 6.67 & 65.8 \\
& IVE & 11.67 & 70.8$\uparrow$ \\
& HSIC & 8.33 & 69.2$\uparrow$ \\
& PASS & \textbf{6.67} & 64.2 \\
& \textbf{AED} & 8.62 & \textbf{63.3} \\
\cmidrule(lr){1-4}
\multirow{5}{*}{DeepEye}
& Baseline & 93.88* & 56.9 \\
& IVE & 93.54* & 60.8$\uparrow$ \\
& \textbf{HSIC} & 93.42* & \textbf{60.0}$\uparrow$ \\
& PASS & 90.66* & 79.2$\uparrow$ \\
& AED & 90.66* & 75.8$\uparrow$ \\
\bottomrule
\end{tabular*}%
}
\end{table}

All suppression methods evaluated here were originally developed
for face or speaker verification, yet they reduce
demographic leakage across every behavioral biometric modality tested.

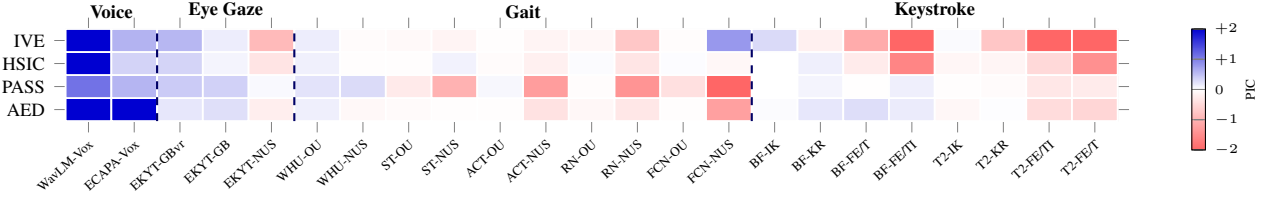
\begin{figure*}[!t]
\centering
\begin{tikzpicture}
\begin{axis}[
  name=hm,
  width=15.5cm, height=2.8cm,
  enlargelimits=false,
  xmin=-0.5, xmax=22.5, ymin=-0.5, ymax=3.5,
  xtick={0,1,2,3,4,5,6,7,8,9,10,11,12,13,14,15,16,17,18,19,20,21,22},
  xticklabels={
    WavLM-Vox, ECAPA-Vox,
    EKYT-GBvr, EKYT-GB, EKYT-NUS,
    WHU-OU, WHU-NUS, ST-OU, ST-NUS, ACT-OU, ACT-NUS, RN-OU, RN-NUS, FCN-OU, FCN-NUS,
    BF-IK, BF-KR, BF-FE/T, {BF-FE/TI},
    T2-IK, T2-KR, {T2-FE/TI}, T2-FE/T,
  },
  xticklabel style={rotate=42, anchor=north east, font=\tiny, inner sep=1.5pt},
  ytick={0,1,2,3},
  yticklabels={AED, PASS, HSIC, IVE},
  yticklabel style={font=\scriptsize, anchor=east},
  colormap={picmap}{
    color(0)=(red!60!white);
    color(0.5)=(white);
    color(1)=(blue!82!black)
  },
  point meta min=-2, point meta max=2,
  colorbar,
  colorbar style={
    xshift=15pt,
    ylabel={PIC},
    ylabel style={font=\tiny, yshift=4pt, xshift=0pt},
    ytick={-2,-1,0,1,2},
    yticklabels={$-2$,$-1$,$0$,$+1$,$+2$},
    yticklabel style={font=\tiny},
    width=0.18cm,
    height=1.6cm,
  },
  grid=none, tick align=outside, clip=false,
  axis line style={draw=gray!40},
]

\addplot[
  matrix plot*,
  point meta=explicit,
  draw=white, line width=0.7pt,
  mesh/cols=23,
]
table[meta=z]{
x y z
0 0 7.23
1 0 2.37
2 0 0.19
3 0 0.25
4 0 -0.22
5 0 0.126
6 0 -0.087
7 0 -0.061
8 0 -0.022
9 0 -0.025
10 0 -0.376
11 0 -0.098
12 0 -0.318
13 0 -0.021
14 0 -1.239
15 0 0.03
16 0 0.19
17 0 0.26
18 0 0.16
19 0 -0.11
20 0 0.02
21 0 -0.46
22 0 -0.54

0 1 1.11
1 1 0.58
2 1 0.41
3 1 0.36
4 1 0.05
5 1 0.224
6 1 0.285
7 1 -0.273
8 1 -1.001
9 1 0.067
10 1 -1.284
11 1 -0.035
12 1 -1.394
13 1 -0.403
14 1 -4.291
15 1 0
16 1 0.09
17 1 -0.01
18 1 0.13
19 1 -0.03
20 1 -0.05
21 1 -0.31
22 1 -0.26

0 2 2.36
1 2 0.35
2 2 0.34
3 2 0.09
4 2 -0.35
5 2 0.122
6 2 -0.021
7 2 -0.02
8 2 0.106
9 2 -0.052
10 2 -0.2
11 2 0.032
12 2 -0.335
13 2 0.021
14 2 -0.11
15 2 0
16 2 0.13
17 2 -0.26
18 2 -1.6
19 2 -0.12
20 2 -0.13
21 2 -0.5
22 2 -1.45

0 3 3.27
1 3 0.6
2 3 0.57
3 3 0.14
4 3 -0.9
5 3 0.143
6 3 -0.053
7 3 -0.077
8 3 -0.141
9 3 -0.027
10 3 -0.148
11 3 -0.107
12 3 -0.725
13 3 -0.04
14 3 0.807
15 3 0.29
16 3 -0.2
17 3 -1.1
18 3 -2.49
19 3 0.04
20 3 -0.74
21 3 -2.58
22 3 -2.24
};

%% Modality group separators
\addplot[blue!40!black, thick, dashed, forget plot]
  coordinates{(1.5,-0.5)(1.5,3.5)};

\addplot[blue!40!black, thick, dashed, forget plot]
  coordinates{(4.5,-0.5)(4.5,3.5)};

\addplot[blue!40!black, thick, dashed, forget plot]
  coordinates{(14.5,-0.5)(14.5,3.5)};

%% Modality group headers
\node[font=\scriptsize\bfseries, anchor=south] at (axis cs:0.5,3.55)
  {Voice};
\node[font=\scriptsize\bfseries, anchor=south] at (axis cs:3.0,3.55)
  {Eye Gaze};
\node[font=\scriptsize\bfseries, anchor=south] at (axis cs:9.5,3.55)
  {Gait};
\node[font=\scriptsize\bfseries, anchor=south] at (axis cs:18.5,3.55)
  {Keystroke};

\end{axis}
\end{tikzpicture}
\caption{PIC heatmap of privacy-utility trade-offs for post-hoc suppression methods across modalities and authentication models. Blue indicates privacy gain exceeds authentication loss; red indicates the opposite.}
\label{fig:heatmap}
\end{figure*}

\textbf{Gaze.} ~\Cref{tab:all_results} shows that all four
post-hoc methods reduce leakage on GazeBase EKYT, with PASS
achieving the largest post-hoc reduction in that setting.
On GazeBaseVR, three of four methods simultaneously improve
both privacy and EER. On DeepEyedentification, HSIC drives
leakage to near-chance (51.9\%). The gaze columns ~\Cref{fig:heatmap} are predominantly blue,
confirming positive privacy--utility trade-off across configurations.

\textbf{Voice.} The strongest evidence comes from ~\Cref{tab:voice}: suppressors reduce ECAPA-TDNN gender from 86.1\% to 52.1\%
with minimal effect on EER and WavLM from 94.8\% to
57.0\%, recovering 94\% of the theoretical
ceiling. The voice columns in ~\Cref{fig:heatmap} are the deepest blue in the study,
and the voice suppression frontier in ~\Cref{fig:pvi_all_modalities} descends further than
any other modality.

\textbf{Keystroke.} ~\Cref{tab:keystroke_all} shows PASS
and AED reduce leakage across BehaveFormer and Type2Branch on
many datasets.

\textbf{Gait.} ~\Cref{tab:gait_all} shows AED reduces
NUS-IGD FG-FCN gender leakage from 70.4\% to 62.0\%, and HSIC reduces it to 58.0\%. These reductions
demonstrate that methods developed outside the gait domain
can locate and remove demographic structure from IMU-based
identity embeddings.

This shows that demographic
structure holds geometrically consistent subspaces across modalities and suppression methods
are transferable beyond their development domain, subject to per-modality validation.

\begin{table}[!t]
\centering
\caption{Gait Demographic Suppression; “–” indicates not reported; “↑” indicates increased leakage}
\label{tab:gait_all}

\small
\setlength{\tabcolsep}{2pt}
\renewcommand{\arraystretch}{0.84}

\resizebox{\columnwidth}{!}{%
\begin{tabular*}{\columnwidth}{@{\extracolsep{\fill}}l l r r r@{}}
\toprule
\textbf{Model} & \textbf{Method}
& \textbf{EER}
& \textbf{Gen.} & \textbf{Age} \\
\midrule

% ================= OU-ISIR =================
\multicolumn{5}{l}{\textbf{Dataset: OU-ISIR}} \\
\cmidrule(lr){1-5}

\multirow{5}{*}{whuGAIT}
& Baseline & 35.4 & 61.9 & 61.3 \\
& HSIC & 31.1 & 58.3 & 56.5 \\
& AED & 30.9 & 61.6 & 58.6 \\
& \textbf{IVE} & 30.4 & \textbf{56.7} & 52.7 \\
& PASS & 27.5 & 62.8$\uparrow$ & 59.9 \\

\cmidrule(lr){1-5}

\multirow{5}{*}{FG-FCN}
& Baseline & 5.64 & 70.4 & 69.0 \\
& \textbf{HSIC} & 5.53 & \textbf{53.5} & 57.2 \\
& AED & 5.76 & 69.8 & 68.1 \\
& IVE & 5.87 & 66.9 & 58.0 \\
& PASS & 7.91 & 72.8$\uparrow$ & 70.6$\uparrow$ \\

\cmidrule(lr){1-5}

\multirow{5}{*}{FG-RN}
& Baseline & 9.39 & 69.5 & 66.0 \\
& HSIC & 9.10 & 59.4 & 53.2 \\
& \textbf{AED} & 10.32 & \textbf{55.8} & 63.7 \\
& IVE & 10.40 & 68.8 & 56.1 \\
& PASS & 9.72 & 69.2 & 65.6 \\

\cmidrule(lr){1-5}

\multirow{5}{*}{ST}
& Baseline & 16.6 & 63.7 & 55.5 \\
& HSIC & 16.9 & 62.1 & 55.3 \\
& \textbf{AED} & 17.6 & \textbf{51.2} & 56.2$\uparrow$ \\
& IVE & 17.9 & 61.5 & 55.3 \\
& PASS & 21.1 & 64.0$\uparrow$ & 54.8 \\

\cmidrule(lr){1-5}

\multirow{5}{*}{ACT}
& Baseline & 14.9 & 67.3 & 58.2 \\
& \textbf{HSIC} & 15.7 & \textbf{64.5} & 57.7 \\
& AED & 15.3 & 67.0 & 59.6$\uparrow$ \\
& IVE & 15.3 & 68.6$\uparrow$ & 57.1 \\
& PASS & 13.9 & 67.5$\uparrow$ & 56.9 \\

\midrule

% ================= NUS-IGD =================
\multicolumn{5}{l}{\textbf{Dataset: NUS-IGD}} \\
\cmidrule(lr){1-5}

\multirow{5}{*}{whuGAIT}
& Baseline & 24.3 & 50.2 & -- \\
& \textbf{HSIC} & 24.8 & \textbf{50.8}$\uparrow$ & -- \\
& AED & 26.4 & 50.4 & -- \\
& IVE & 25.5 & 57.0$\uparrow$ & -- \\
& PASS & 17.4 & 51.2$\uparrow$ & -- \\

\cmidrule(lr){1-5}

\multirow{5}{*}{FG-FCN}
& Baseline & 0.62 & 70.4 & -- \\
& HSIC & 0.69 & 58.0 & -- \\
& \textbf{AED} & 1.39 & \textbf{62.0} & -- \\
& IVE & 0.12 & 63.8 & -- \\
& PASS & 3.28 & 73.6$\uparrow$ & -- \\

\cmidrule(lr){1-5}

\multirow{5}{*}{FG-RN}
& Baseline & 2.08 & 56.0 & -- \\
& \textbf{HSIC} & 2.78 & \textbf{59.5}$\uparrow$ & -- \\
& AED & 2.74 & 59.9$\uparrow$ & -- \\
& IVE & 3.59 & 50.2 & -- \\
& PASS & 4.98 & 55.0 & -- \\

\cmidrule(lr){1-5}

\multirow{5}{*}{ST}
& Baseline & 5.44 & 50.6 & -- \\
& HSIC & 4.86 & 53.8$\uparrow$ & -- \\
& AED & 5.56 & 54.8$\uparrow$ & -- \\
& \textbf{IVE} & 6.21 & \textbf{55.4}$\uparrow$ & -- \\
& PASS & 10.88 & 57.7$\uparrow$ & -- \\

\cmidrule(lr){1-5}

\multirow{5}{*}{ACT}
& Baseline & 4.69 & 67.1 & -- \\
& HSIC & 5.63 & 65.0 & -- \\
& \textbf{AED} & 6.46 & \textbf{54.4} & -- \\
& IVE & 5.38 & 72.4$\uparrow$ & -- \\
& PASS & 10.71 & 71.0$\uparrow$ & -- \\

\bottomrule
\end{tabular*}%
}%
\end{table}

% Heatmap

%--------------------------------------------------------------------
% ROC / DET figure — GazeBase EKYT, all suppression methods
%--------------------------------------------------------------------
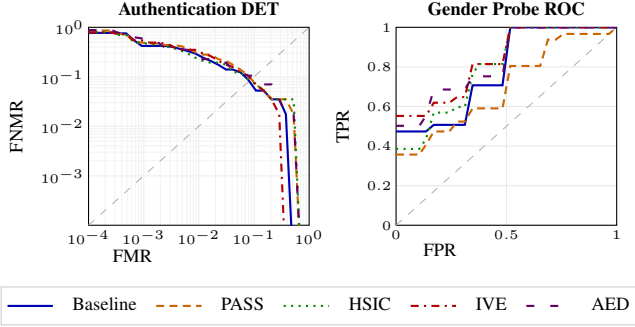
\begin{figure}[t]
\centering
\begin{tikzpicture}
\begin{groupplot}[
  group style={
    group size=2 by 1,
    horizontal sep=1.15cm,
  },
  width=4.5cm, height=4.2cm,
  tick align=outside,
  tick pos=left,
  title style={yshift=-5pt}, tick style={draw=none},
  minor tick num=0,
 legend style={
  font=\scriptsize,
  draw=gray!40,
  inner sep=2pt,
  /tikz/column sep=0.5em,
  row sep=0.1em
},
legend columns=5,
]

%%──── Panel 1: Authentication DET ────────────────────────────────────
\nextgroupplot[
  xmode=log, ymode=log,
  xmin=1e-4, xmax=1,
  ymin=1e-4, ymax=1,
  xlabel={FMR},
  ylabel={FNMR},
  xlabel style={font=\scriptsize, yshift=5pt, xshift=-25pt},
  ylabel style={font=\scriptsize},
  title={\scriptsize\bfseries Authentication DET},
  title style={font=\scriptsize\bfseries},
  xtick={1e-4,1e-3,1e-2,1e-1,1},
  ytick={1e-3,1e-2,1e-1,1},
  xticklabel style={font=\tiny, yshift=6pt},
  yticklabel style={font=\tiny, xshift=6pt},
  grid=both,
  grid style={gray!10, very thin},
  legend to name=gazeallleg,
]
\addplot[gray!60, dashed, thin, forget plot] coordinates {(1e-4,1e-4)(1,1)};
\addplot[blue!70!black,   thick, solid] table[x=fmr,y=fnmr]{
fmr fnmr
0.00010 0.77193
0.00014 0.77193
0.00019 0.77193
0.00026 0.77193
0.00035 0.75439
0.00048 0.75439
0.00067 0.54386
0.00092 0.42105
0.00128 0.42105
0.00176 0.42105
0.00244 0.42105
0.00322 0.40351
0.00445 0.36842
0.00615 0.33333
0.00850 0.29825
0.01176 0.24561
0.01626 0.21053
0.02248 0.17544
0.03108 0.14035
0.04103 0.14035
0.05672 0.12281
0.07843 0.08772
0.10844 0.05263
0.14993 0.05263
0.20729 0.03509
0.28661 0.03509
0.37835 0.01754
0.52311 0.00001
0.72326 0.00001
1.00000 0.00001
};
\addlegendentry{Baseline}
\addplot[orange!80!black, thick, densely dashed] table[x=fmr,y=fnmr]{
fmr fnmr
0.00010 0.85965
0.00014 0.85965
0.00019 0.85965
0.00026 0.85965
0.00035 0.73684
0.00048 0.73684
0.00067 0.49123
0.00092 0.49123
0.00128 0.47368
0.00176 0.47368
0.00244 0.43860
0.00322 0.43860
0.00445 0.42105
0.00615 0.38596
0.00850 0.31579
0.01176 0.28070
0.01626 0.28070
0.02248 0.22807
0.03108 0.21053
0.04103 0.17544
0.05672 0.14035
0.07843 0.10526
0.10844 0.07018
0.14993 0.05263
0.20729 0.03509
0.28661 0.03509
0.37835 0.03509
0.52311 0.01754
0.72326 0.00001
1.00000 0.00001
};
\addlegendentry{PASS}
\addplot[green!55!black,  thick, dotted] table[x=fmr,y=fnmr]{
fmr fnmr
0.00010 0.78947
0.00014 0.78947
0.00019 0.78947
0.00026 0.78947
0.00035 0.70175
0.00048 0.70175
0.00067 0.50877
0.00092 0.49123
0.00128 0.45614
0.00176 0.43860
0.00244 0.42105
0.00322 0.38596
0.00445 0.35088
0.00615 0.29825
0.00850 0.24561
0.01176 0.21053
0.01626 0.19298
0.02248 0.19298
0.03108 0.17544
0.04103 0.12281
0.05672 0.12281
0.07843 0.08772
0.10844 0.07018
0.14993 0.05263
0.20729 0.03509
0.28661 0.03509
0.37835 0.03509
0.52311 0.03509
0.72326 0.00001
1.00000 0.00001
};
\addlegendentry{HSIC}
\addplot[red!70!black,    thick, dash dot] table[x=fmr,y=fnmr]{
fmr fnmr
0.00010 0.78947
0.00014 0.78947
0.00019 0.78947
0.00026 0.78947
0.00035 0.71930
0.00048 0.71930
0.00067 0.54386
0.00092 0.54386
0.00128 0.49123
0.00176 0.49123
0.00244 0.47368
0.00322 0.42105
0.00445 0.35088
0.00615 0.35088
0.00850 0.35088
0.01176 0.29825
0.01626 0.26316
0.02248 0.21053
0.03108 0.17544
0.04103 0.15789
0.05672 0.12281
0.07843 0.10526
0.10844 0.07018
0.14993 0.05263
0.20729 0.03509
0.28661 0.01754
0.37835 0.00001
0.52311 0.00001
0.72326 0.00001
1.00000 0.00001
};
\addlegendentry{IVE}
\addplot[violet!80!black, thick, loosely dashed] table[x=fmr,y=fnmr]{
fmr fnmr
0.00010 0.87719
0.00014 0.87719
0.00019 0.87719
0.00026 0.87719
0.00035 0.77193
0.00048 0.77193
0.00067 0.63158
0.00092 0.59649
0.00128 0.56140
0.00176 0.50877
0.00244 0.38596
0.00322 0.36842
0.00445 0.36842
0.00615 0.35088
0.00850 0.29825
0.01176 0.22807
0.01626 0.22807
0.02248 0.19298
0.03108 0.19298
0.04103 0.14035
0.05672 0.12281
0.07843 0.10526
0.10844 0.10526
0.14993 0.07018
0.20729 0.07018
0.28661 0.03509
0.37835 0.01754
0.52311 0.01754
0.72326 0.00001
1.00000 0.00001
};
\addlegendentry{AED}

%%──── Panel 2: Gender Probe ROC ─────────────────────────────────────
\nextgroupplot[
  xmin=0, xmax=1, ymin=0, ymax=1,
  xlabel={FPR},
  ylabel={TPR},
  xlabel style={font=\scriptsize, yshift=5pt, xshift=-25pt},
  ylabel style={font=\scriptsize},
  title={\scriptsize\bfseries Gender Probe ROC},
  title style={font=\scriptsize\bfseries},
  xtick={0,0.5,1.0},
  ytick={0,0.2,0.4,0.6,0.8,1.0},
  xticklabel style={font=\tiny, yshift=6pt},
  yticklabel style={font=\tiny, xshift=6pt},
  grid=both,
  grid style={gray!20, very thin},
]
\addplot[gray!60, dashed, thin, forget plot] coordinates {(0,0)(1,1)};
\addplot[blue!70!black,   thick, solid, forget plot] table[x=fpr,y=tpr]{
fpr tpr
0.00000 0.47381
0.03518 0.47381
0.07035 0.47381
0.10553 0.47381
0.13568 0.47381
0.17085 0.50714
0.20603 0.50714
0.24121 0.50714
0.27638 0.50714
0.31156 0.50714
0.34673 0.70714
0.37688 0.70714
0.41206 0.70714
0.44724 0.70714
0.48241 0.70714
0.51759 1.00000
0.55276 1.00000
0.58794 1.00000
0.62312 1.00000
0.65327 1.00000
0.68844 1.00000
0.72362 1.00000
0.75879 1.00000
0.79397 1.00000
0.82915 1.00000
0.86432 1.00000
0.89447 1.00000
0.92965 1.00000
0.96482 1.00000
1.00000 1.00000
};
\addplot[orange!80!black, thick, densely dashed, forget plot] table[x=fpr,y=tpr]{
fpr tpr
0.00000 0.35714
0.03518 0.35714
0.07035 0.35714
0.10553 0.35714
0.13568 0.40714
0.17085 0.47381
0.20603 0.47381
0.24121 0.47381
0.27638 0.52381
0.31156 0.52381
0.34673 0.59048
0.37688 0.59048
0.41206 0.59048
0.44724 0.59048
0.48241 0.59048
0.51759 0.80476
0.55276 0.80476
0.58794 0.80476
0.62312 0.80476
0.65327 0.80476
0.68844 0.93810
0.72362 0.93810
0.75879 0.96667
0.79397 0.96667
0.82915 0.96667
0.86432 0.96667
0.89447 0.96667
0.92965 0.96667
0.96482 0.96667
1.00000 1.00000
};
\addplot[green!55!black,  thick, dotted, forget plot] table[x=fpr,y=tpr]{
fpr tpr
0.00000 0.38571
0.03518 0.38571
0.07035 0.38571
0.10553 0.38571
0.13568 0.43571
0.17085 0.56905
0.20603 0.56905
0.24121 0.56905
0.27638 0.59762
0.31156 0.59762
0.34673 0.76429
0.37688 0.81429
0.41206 0.81429
0.44724 0.81429
0.48241 0.81429
0.51759 1.00000
0.55276 1.00000
0.58794 1.00000
0.62312 1.00000
0.65327 1.00000
0.68844 1.00000
0.72362 1.00000
0.75879 1.00000
0.79397 1.00000
0.82915 1.00000
0.86432 1.00000
0.89447 1.00000
0.92965 1.00000
0.96482 1.00000
1.00000 1.00000
};
\addplot[red!70!black,    thick, dash dot, forget plot] table[x=fpr,y=tpr]{
fpr tpr
0.00000 0.55238
0.03518 0.55238
0.07035 0.55238
0.10553 0.55238
0.13568 0.55238
0.17085 0.61905
0.20603 0.61905
0.24121 0.61905
0.27638 0.64762
0.31156 0.64762
0.34673 0.81429
0.37688 0.81429
0.41206 0.81429
0.44724 0.81429
0.48241 0.81429
0.51759 1.00000
0.55276 1.00000
0.58794 1.00000
0.62312 1.00000
0.65327 1.00000
0.68844 1.00000
0.72362 1.00000
0.75879 1.00000
0.79397 1.00000
0.82915 1.00000
0.86432 1.00000
0.89447 1.00000
0.92965 1.00000
0.96482 1.00000
1.00000 1.00000
};
\addplot[violet!80!black, thick, loosely dashed, forget plot] table[x=fpr,y=tpr]{
fpr tpr
0.00000 0.50238
0.03518 0.50238
0.07035 0.50238
0.10553 0.50238
0.13568 0.55238
0.17085 0.68571
0.20603 0.68571
0.24121 0.68571
0.27638 0.68571
0.31156 0.68571
0.34673 0.75238
0.37688 0.75238
0.41206 0.75238
0.44724 0.75238
0.48241 0.75238
0.51759 1.00000
0.55276 1.00000
0.58794 1.00000
0.62312 1.00000
0.65327 1.00000
0.68844 1.00000
0.72362 1.00000
0.75879 1.00000
0.79397 1.00000
0.82915 1.00000
0.86432 1.00000
0.89447 1.00000
0.92965 1.00000
0.96482 1.00000
1.00000 1.00000
};

\end{groupplot}
\end{tikzpicture}
\par\vspace{5pt}
\pgfplotslegendfromname{gazeallleg}
\caption{Authentication DET and gender-probe ROC for GazeBase EKYT under baseline and suppression settings.}
\label{fig:roc}
\end{figure}

%--------------------------------------------------------------------
%--------------------------------------------------------------------
% Keystroke ROC — BehaveFormer and Type2Branch, Touch vs Touch+IMU
%--------------------------------------------------------------------
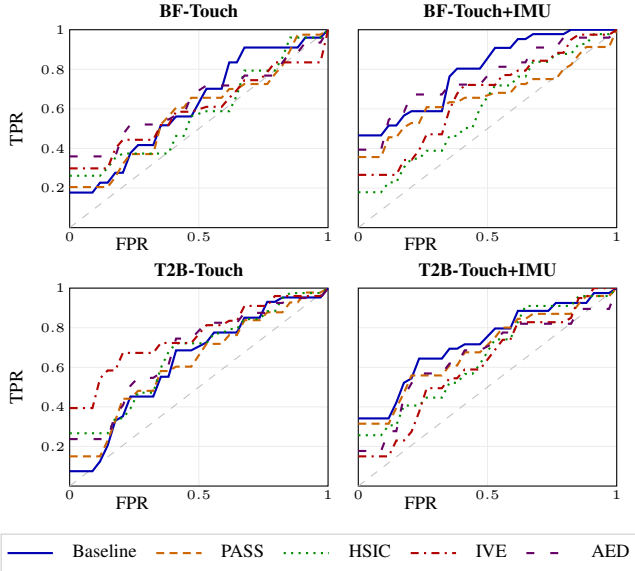
\begin{figure}[!t]
\centering
% \vspace{0.75\baselineskip}
\begin{tikzpicture}
\begin{groupplot}[
  group style={
    group size=2 by 2,
    horizontal sep=0.40cm,
    vertical sep=0.8cm,
  },
  width=5cm, height=4.2cm,
  xmin=0, xmax=1, ymin=0, ymax=1,
  xlabel={FPR}, ylabel={TPR},
  xlabel style={font=\scriptsize, yshift=8pt, xshift=-25pt}, ylabel style={font=\scriptsize},
  title style={yshift=-5pt},
  xtick={0,0.5,1.0},
  ytick={0.2,0.4,0.6,0.8,1.0},
  xticklabel style={font=\tiny, yshift=6pt}, yticklabel style={font=\tiny, xshift=6pt}, 
  tick align=outside, tick pos=left,tick style={draw=none},
  grid=both, grid style={gray!15, very thin},
legend style={
  font=\scriptsize,
  draw=gray!40,
  inner sep=2pt,
  /tikz/column sep=0.5em,
  row sep=0.2em
},
legend columns=5,
]

%% ────────────── Panel A: BehaveFormer — Touch ───────────────────────
\nextgroupplot[
  title={\scriptsize\bfseries BF-Touch},
  legend entries={Baseline, PASS, HSIC, IVE, AED},
  legend to name=kslegend,
]
\addplot[gray!50, dashed, very thin, forget plot] coordinates{(0,0)(1,1)};
\addplot[blue!70!black,   thick,  solid]         table[x=fpr,y=tpr]{
fpr tpr
0.0000 0.1772
0.0294 0.1772
0.0588 0.1772
0.0882 0.1772
0.1176 0.2272
0.1471 0.2272
0.1765 0.2772
0.2059 0.2772
0.2353 0.3772
0.2647 0.4172
0.2941 0.4172
0.3235 0.4172
0.3529 0.5172
0.3824 0.5172
0.4118 0.5617
0.4412 0.5617
0.4706 0.5617
0.5000 0.6317
0.5294 0.7017
0.5588 0.7017
0.5882 0.7017
0.6176 0.8350
0.6471 0.8350
0.6765 0.9100
0.7059 0.9100
0.7353 0.9100
0.7647 0.9100
0.7941 0.9100
0.8235 0.9100
0.8529 0.9100
0.8824 0.9100
0.9118 0.9600
0.9412 0.9600
0.9706 0.9600
1.0000 1.0000
};
\addplot[orange!80!black, thick,  densely dashed] table[x=fpr,y=tpr]{
fpr tpr
0.0000 0.2050
0.0294 0.2050
0.0588 0.2050
0.0882 0.2050
0.1176 0.2050
0.1471 0.2050
0.1765 0.2550
0.2059 0.3217
0.2353 0.3717
0.2647 0.3717
0.2941 0.3717
0.3235 0.3717
0.3529 0.5217
0.3824 0.5617
0.4118 0.6061
0.4412 0.6061
0.4706 0.6561
0.5000 0.6561
0.5294 0.6561
0.5588 0.6561
0.5882 0.6561
0.6176 0.7006
0.6471 0.7006
0.6765 0.7256
0.7059 0.7256
0.7353 0.7256
0.7647 0.7256
0.7941 0.7756
0.8235 0.8200
0.8529 0.8950
0.8824 0.9750
0.9118 0.9750
0.9412 0.9750
0.9706 0.9750
1.0000 1.0000
};
\addplot[green!55!black,  thick,  dotted]         table[x=fpr,y=tpr]{
fpr tpr
0.0000 0.2622
0.0294 0.2622
0.0588 0.2622
0.0882 0.2622
0.1176 0.2622
0.1471 0.3022
0.1765 0.3522
0.2059 0.3744
0.2353 0.3744
0.2647 0.3744
0.2941 0.3744
0.3235 0.3744
0.3529 0.3744
0.3824 0.3744
0.4118 0.4633
0.4412 0.4633
0.4706 0.5633
0.5000 0.5758
0.5294 0.5883
0.5588 0.5883
0.5882 0.5883
0.6176 0.5883
0.6471 0.6683
0.6765 0.7933
0.7059 0.7933
0.7353 0.7933
0.7647 0.7933
0.7941 0.7933
0.8235 0.8600
0.8529 0.9600
0.8824 0.9600
0.9118 0.9600
0.9412 0.9600
0.9706 0.9600
1.0000 1.0000
};
\addplot[red!70!black,    thick,  dash dot]        table[x=fpr,y=tpr]{
fpr tpr
0.0000 0.2994
0.0294 0.2994
0.0588 0.2994
0.0882 0.2994
0.1176 0.2994
0.1471 0.2994
0.1765 0.3994
0.2059 0.4439
0.2353 0.4439
0.2647 0.4439
0.2941 0.4439
0.3235 0.4439
0.3529 0.5189
0.3824 0.5189
0.4118 0.5856
0.4412 0.5856
0.4706 0.5856
0.5000 0.5981
0.5294 0.6106
0.5588 0.6106
0.5882 0.6106
0.6176 0.6550
0.6471 0.6950
0.6765 0.7450
0.7059 0.7450
0.7353 0.7450
0.7647 0.7850
0.7941 0.8350
0.8235 0.8350
0.8529 0.8350
0.8824 0.8350
0.9118 0.8350
0.9412 0.8350
0.9706 0.8350
1.0000 1.0000
};
\addplot[violet!80!black, thick,  loosely dashed]  table[x=fpr,y=tpr]{
fpr tpr
0.0000 0.3600
0.0294 0.3600
0.0588 0.3600
0.0882 0.3600
0.1176 0.3600
0.1471 0.3600
0.1765 0.3600
0.2059 0.4711
0.2353 0.5211
0.2647 0.5211
0.2941 0.5211
0.3235 0.5211
0.3529 0.5461
0.3824 0.5461
0.4118 0.5683
0.4412 0.5683
0.4706 0.6683
0.5000 0.6933
0.5294 0.7183
0.5588 0.7183
0.5882 0.7183
0.6176 0.7183
0.6471 0.7183
0.6765 0.7683
0.7059 0.7683
0.7353 0.7683
0.7647 0.7683
0.7941 0.7683
0.8235 0.8350
0.8529 0.8850
0.8824 0.8850
0.9118 0.9350
0.9412 0.9350
0.9706 0.9350
1.0000 1.0000
};
\node[font=\tiny, anchor=north, align=center, text width=3.2cm]
  at ([yshift=-13pt]current axis.south)
  {AUC:~Baseline\,0.62~~PASS\,0.59~~HSIC\,0.58\\IVE\,0.59~~AED\,0.64};

%% ────────────── Panel B: BehaveFormer — Touch+IMU ───────────────────
\nextgroupplot[
  title={\scriptsize\bfseries BF-Touch+IMU},
  ylabel={}, yticklabels={},
]
\addplot[gray!50, dashed, very thin, forget plot] coordinates{(0,0)(1,1)};
\addplot[blue!70!black,   thick,  solid,          forget plot] table[x=fpr,y=tpr]{
fpr tpr
0.0000 0.4661
0.0294 0.4661
0.0588 0.4661
0.0882 0.4661
0.1176 0.5161
0.1471 0.5161
0.1765 0.5661
0.2059 0.5883
0.2353 0.5883
0.2647 0.5883
0.2941 0.5883
0.3235 0.5883
0.3529 0.7633
0.3824 0.8033
0.4118 0.8033
0.4412 0.8033
0.4706 0.8033
0.5000 0.8558
0.5294 0.9083
0.5588 0.9083
0.5882 0.9083
0.6176 0.9528
0.6471 0.9528
0.6765 0.9778
0.7059 0.9778
0.7353 0.9778
0.7647 0.9778
0.7941 0.9778
0.8235 1.0000
0.8529 1.0000
0.8824 1.0000
0.9118 1.0000
0.9412 1.0000
0.9706 1.0000
1.0000 1.0000
};
\addplot[orange!80!black, thick,  densely dashed,  forget plot] table[x=fpr,y=tpr]{
fpr tpr
0.0000 0.3567
0.0294 0.3567
0.0588 0.3567
0.0882 0.3567
0.1176 0.4567
0.1471 0.4567
0.1765 0.5067
0.2059 0.5289
0.2353 0.5289
0.2647 0.6089
0.2941 0.6089
0.3235 0.6089
0.3529 0.6339
0.3824 0.6339
0.4118 0.6561
0.4412 0.6561
0.4706 0.6561
0.5000 0.6686
0.5294 0.6811
0.5588 0.6811
0.5882 0.6811
0.6176 0.7256
0.6471 0.7256
0.6765 0.7506
0.7059 0.7506
0.7353 0.7506
0.7647 0.7506
0.7941 0.8006
0.8235 0.8228
0.8529 0.8728
0.8824 0.9128
0.9118 0.9128
0.9412 0.9128
0.9706 0.9128
1.0000 1.0000
};
\addplot[green!55!black,  thick,  dotted,          forget plot] table[x=fpr,y=tpr]{
fpr tpr
0.0000 0.1789
0.0294 0.1789
0.0588 0.1789
0.0882 0.1789
0.1176 0.2289
0.1471 0.2289
0.1765 0.3039
0.2059 0.3483
0.2353 0.3483
0.2647 0.3883
0.2941 0.3883
0.3235 0.3883
0.3529 0.4633
0.3824 0.4633
0.4118 0.5033
0.4412 0.5033
0.4706 0.6033
0.5000 0.6608
0.5294 0.7183
0.5588 0.7183
0.5882 0.7183
0.6176 0.7628
0.6471 0.7628
0.6765 0.8378
0.7059 0.8378
0.7353 0.8378
0.7647 0.8778
0.7941 0.8778
0.8235 0.8778
0.8529 0.9278
0.8824 0.9278
0.9118 0.9778
0.9412 0.9778
0.9706 0.9778
1.0000 1.0000
};
\addplot[red!70!black,    thick,  dash dot,         forget plot] table[x=fpr,y=tpr]{
fpr tpr
0.0000 0.2667
0.0294 0.2667
0.0588 0.2667
0.0882 0.2667
0.1176 0.2667
0.1471 0.2667
0.1765 0.3417
0.2059 0.3417
0.2353 0.3917
0.2647 0.4717
0.2941 0.4717
0.3235 0.4717
0.3529 0.5967
0.3824 0.6767
0.4118 0.7211
0.4412 0.7211
0.4706 0.7211
0.5000 0.7211
0.5294 0.7211
0.5588 0.7711
0.5882 0.7711
0.6176 0.7933
0.6471 0.7933
0.6765 0.8433
0.7059 0.8433
0.7353 0.8433
0.7647 0.8833
0.7941 0.8833
0.8235 0.9500
0.8529 0.9750
0.8824 0.9750
0.9118 0.9750
0.9412 0.9750
0.9706 0.9750
1.0000 1.0000
};
\addplot[violet!80!black, thick,  loosely dashed,   forget plot] table[x=fpr,y=tpr]{
fpr tpr
0.0000 0.3939
0.0294 0.3939
0.0588 0.3939
0.0882 0.3939
0.1176 0.5439
0.1471 0.5839
0.1765 0.5839
0.2059 0.6728
0.2353 0.6728
0.2647 0.6728
0.2941 0.6728
0.3235 0.6728
0.3529 0.7228
0.3824 0.7228
0.4118 0.7228
0.4412 0.7228
0.4706 0.7228
0.5000 0.7678
0.5294 0.8128
0.5588 0.8128
0.5882 0.8128
0.6176 0.8350
0.6471 0.8350
0.6765 0.9100
0.7059 0.9100
0.7353 0.9100
0.7647 0.9100
0.7941 0.9600
0.8235 0.9600
0.8529 0.9600
0.8824 0.9600
0.9118 0.9600
0.9412 0.9600
0.9706 0.9600
1.0000 1.0000
};
\node[font=\tiny, anchor=north, align=center, text width=3.2cm]
  at ([yshift=-13pt]current axis.south)
  {AUC:~Baseline\,0.75~~PASS\,0.63~~HSIC\,0.58\\IVE\,0.66~~AED\,0.70};

%% ────────────── Panel C: Type2Branch — Touch ────────────────────────
\nextgroupplot[
  title={\scriptsize\bfseries T2B-Touch},
]
\addplot[gray!50, dashed, very thin, forget plot] coordinates{(0,0)(1,1)};
\addplot[blue!70!black,   thick,  solid,          forget plot] table[x=fpr,y=tpr]{
fpr tpr
0.0000 0.0750
0.0294 0.0750
0.0588 0.0750
0.0882 0.0750
0.1176 0.1250
0.1471 0.2050
0.1765 0.3300
0.2059 0.3522
0.2353 0.4522
0.2647 0.4522
0.2941 0.4522
0.3235 0.4522
0.3529 0.5522
0.3824 0.5522
0.4118 0.6856
0.4412 0.6856
0.4706 0.6856
0.5000 0.7056
0.5294 0.7256
0.5588 0.7756
0.5882 0.7756
0.6176 0.7756
0.6471 0.7756
0.6765 0.8506
0.7059 0.8506
0.7353 0.8506
0.7647 0.9306
0.7941 0.9306
0.8235 0.9528
0.8529 0.9528
0.8824 0.9528
0.9118 0.9528
0.9412 0.9528
0.9706 0.9528
1.0000 1.0000
};
\addplot[orange!80!black, thick,  densely dashed,  forget plot] table[x=fpr,y=tpr]{
fpr tpr
0.0000 0.1500
0.0294 0.1500
0.0588 0.1500
0.0882 0.1500
0.1176 0.1500
0.1471 0.2300
0.1765 0.3300
0.2059 0.4411
0.2353 0.4411
0.2647 0.4811
0.2941 0.4811
0.3235 0.4811
0.3529 0.5811
0.3824 0.5811
0.4118 0.6033
0.4412 0.6033
0.4706 0.6033
0.5000 0.6608
0.5294 0.7183
0.5588 0.7183
0.5882 0.7183
0.6176 0.7628
0.6471 0.7628
0.6765 0.8378
0.7059 0.8378
0.7353 0.8378
0.7647 0.8778
0.7941 0.8778
0.8235 0.8778
0.8529 0.9278
0.8824 0.9278
0.9118 0.9778
0.9412 0.9778
0.9706 0.9778
1.0000 1.0000
};
\addplot[green!55!black,  thick,  dotted,          forget plot] table[x=fpr,y=tpr]{
fpr tpr
0.0000 0.2667
0.0294 0.2667
0.0588 0.2667
0.0882 0.2667
0.1176 0.2667
0.1471 0.2667
0.1765 0.3417
0.2059 0.3417
0.2353 0.3917
0.2647 0.4717
0.2941 0.4717
0.3235 0.4717
0.3529 0.5967
0.3824 0.6767
0.4118 0.7211
0.4412 0.7211
0.4706 0.7211
0.5000 0.7211
0.5294 0.7211
0.5588 0.7711
0.5882 0.7711
0.6176 0.7933
0.6471 0.7933
0.6765 0.8433
0.7059 0.8433
0.7353 0.8433
0.7647 0.8833
0.7941 0.8833
0.8235 0.9500
0.8529 0.9750
0.8824 0.9750
0.9118 0.9750
0.9412 0.9750
0.9706 0.9750
1.0000 1.0000
};
\addplot[red!70!black,    thick,  dash dot,         forget plot] table[x=fpr,y=tpr]{
fpr tpr
0.0000 0.3939
0.0294 0.3939
0.0588 0.3939
0.0882 0.3939
0.1176 0.5439
0.1471 0.5839
0.1765 0.5839
0.2059 0.6728
0.2353 0.6728
0.2647 0.6728
0.2941 0.6728
0.3235 0.6728
0.3529 0.7228
0.3824 0.7228
0.4118 0.7228
0.4412 0.7228
0.4706 0.7228
0.5000 0.7678
0.5294 0.8128
0.5588 0.8128
0.5882 0.8128
0.6176 0.8350
0.6471 0.8350
0.6765 0.9100
0.7059 0.9100
0.7353 0.9100
0.7647 0.9100
0.7941 0.9600
0.8235 0.9600
0.8529 0.9600
0.8824 0.9600
0.9118 0.9600
0.9412 0.9600
0.9706 0.9600
1.0000 1.0000
};
\addplot[violet!80!black, thick,  loosely dashed,   forget plot] table[x=fpr,y=tpr]{
fpr tpr
0.0000 0.2372
0.0294 0.2372
0.0588 0.2372
0.0882 0.2372
0.1176 0.2372
0.1471 0.2372
0.1765 0.3622
0.2059 0.4067
0.2353 0.5067
0.2647 0.5467
0.2941 0.5467
0.3235 0.5467
0.3529 0.5717
0.3824 0.6117
0.4118 0.7450
0.4412 0.7450
0.4706 0.7450
0.5000 0.7850
0.5294 0.8250
0.5588 0.8250
0.5882 0.8250
0.6176 0.8250
0.6471 0.8250
0.6765 0.8750
0.7059 0.8750
0.7353 0.8750
0.7647 0.8750
0.7941 0.9250
0.8235 0.9250
0.8529 0.9500
0.8824 0.9500
0.9118 0.9500
0.9412 0.9500
0.9706 0.9500
1.0000 1.0000
};
\node[font=\tiny, anchor=north, align=center, text width=3.2cm]
  at ([yshift=-13pt]current axis.south)
  {AUC:~Baseline\,0.63~~PASS\,0.63~~HSIC\,0.66\\IVE\,0.76~~AED\,0.68};

%% ────────────── Panel D: Type2Branch — Touch+IMU ────────────────────
\nextgroupplot[
  title={\scriptsize\bfseries T2B-Touch+IMU},
  ylabel={}, yticklabels={},
]
\addplot[gray!50, dashed, very thin, forget plot] coordinates{(0,0)(1,1)};
\addplot[blue!70!black,   thick,  solid,          forget plot] table[x=fpr,y=tpr]{
fpr tpr
0.0000 0.3417
0.0294 0.3417
0.0588 0.3417
0.0882 0.3417
0.1176 0.3417
0.1471 0.4217
0.1765 0.5217
0.2059 0.5439
0.2353 0.6439
0.2647 0.6439
0.2941 0.6439
0.3235 0.6439
0.3529 0.6939
0.3824 0.6939
0.4118 0.7161
0.4412 0.7161
0.4706 0.7161
0.5000 0.7561
0.5294 0.7961
0.5588 0.7961
0.5882 0.7961
0.6176 0.8850
0.6471 0.8850
0.6765 0.8850
0.7059 0.8850
0.7353 0.8850
0.7647 0.9250
0.7941 0.9250
0.8235 0.9250
0.8529 0.9250
0.8824 0.9250
0.9118 0.9750
0.9412 0.9750
0.9706 0.9750
1.0000 1.0000
};
\addplot[orange!80!black, thick,  densely dashed,  forget plot] table[x=fpr,y=tpr]{
fpr tpr
0.0000 0.3150
0.0294 0.3150
0.0588 0.3150
0.0882 0.3150
0.1176 0.3150
0.1471 0.3950
0.1765 0.4700
0.2059 0.5589
0.2353 0.5589
0.2647 0.5589
0.2941 0.5589
0.3235 0.5589
0.3529 0.6089
0.3824 0.6089
0.4118 0.6756
0.4412 0.6756
0.4706 0.6756
0.5000 0.7131
0.5294 0.7506
0.5588 0.8006
0.5882 0.8006
0.6176 0.8450
0.6471 0.8450
0.6765 0.8700
0.7059 0.8700
0.7353 0.8700
0.7647 0.8700
0.7941 0.8700
0.8235 0.8700
0.8529 0.9200
0.8824 0.9600
0.9118 0.9600
0.9412 0.9600
0.9706 0.9600
1.0000 1.0000
};
\addplot[green!55!black,  thick,  dotted,          forget plot] table[x=fpr,y=tpr]{
fpr tpr
0.0000 0.2567
0.0294 0.2567
0.0588 0.2567
0.0882 0.2567
0.1176 0.3067
0.1471 0.3067
0.1765 0.4067
0.2059 0.4067
0.2353 0.4067
0.2647 0.4467
0.2941 0.4467
0.3235 0.4467
0.3529 0.5217
0.3824 0.5217
0.4118 0.5661
0.4412 0.5661
0.4706 0.6161
0.5000 0.6786
0.5294 0.7411
0.5588 0.7411
0.5882 0.7411
0.6176 0.8300
0.6471 0.9100
0.6765 0.9100
0.7059 0.9100
0.7353 0.9100
0.7647 0.9100
0.7941 0.9100
0.8235 0.9100
0.8529 0.9600
0.8824 0.9600
0.9118 0.9600
0.9412 0.9600
0.9706 0.9600
1.0000 1.0000
};
\addplot[red!70!black,    thick,  dash dot,         forget plot] table[x=fpr,y=tpr]{
fpr tpr
0.0000 0.1500
0.0294 0.1500
0.0588 0.1500
0.0882 0.1500
0.1176 0.1500
0.1471 0.2300
0.1765 0.2300
0.2059 0.2744
0.2353 0.3744
0.2647 0.4944
0.2941 0.4944
0.3235 0.4944
0.3529 0.5444
0.3824 0.5444
0.4118 0.5889
0.4412 0.5889
0.4706 0.5889
0.5000 0.6389
0.5294 0.6889
0.5588 0.7389
0.5882 0.7389
0.6176 0.8278
0.6471 0.8278
0.6765 0.8278
0.7059 0.8278
0.7353 0.8278
0.7647 0.8278
0.7941 0.8278
0.8235 0.8500
0.8529 0.9500
0.8824 0.9500
0.9118 1.0000
0.9412 1.0000
0.9706 1.0000
1.0000 1.0000
};
\addplot[violet!80!black, thick,  loosely dashed,   forget plot] table[x=fpr,y=tpr]{
fpr tpr
0.0000 0.1772
0.0294 0.1772
0.0588 0.1772
0.0882 0.1772
0.1176 0.2772
0.1471 0.2772
0.1765 0.4522
0.2059 0.5189
0.2353 0.5689
0.2647 0.5689
0.2941 0.5689
0.3235 0.5689
0.3529 0.6189
0.3824 0.6189
0.4118 0.6856
0.4412 0.6856
0.4706 0.6856
0.5000 0.7056
0.5294 0.7256
0.5588 0.7756
0.5882 0.7756
0.6176 0.8200
0.6471 0.8200
0.6765 0.8200
0.7059 0.8200
0.7353 0.8200
0.7647 0.8200
0.7941 0.8200
0.8235 0.8200
0.8529 0.8950
0.8824 0.8950
0.9118 0.8950
0.9412 0.8950
0.9706 0.8950
1.0000 1.0000
};
\node[font=\tiny, anchor=north, align=center, text width=3.2cm]
  at ([yshift=-13pt]current axis.south)
  {AUC:~Baseline\,0.73~~PASS\,0.69~~HSIC\,0.66\\IVE\,0.62~~AED\,0.65};

\end{groupplot}
\end{tikzpicture}
\par\vspace{5pt}
\pgfplotslegendfromname{kslegend}

\caption{Gender-probe ROC for FETA keystroke settings BF = BehaveFormer, T2B = Type2Branch}
\label{fig:ks_roc}
\end{figure}

%--------------------------------------------------------------------
\section{Discussion}
%--------------------------------------------------------------------
\textbf{Training objectives routinely entangle identity and demographic structure.}
Every model in ~\Cref{tab:baseline} leaks gender above chance
across CNNs, LSTMs, and Transformers in four modalities and nine
datasets. The PIC heatmap (~\Cref{fig:heatmap}) confirms the
breadth: no column sits at baseline. This suggests that current
training objectives, which optimize identity discriminability without
an explicit privacy constraint, routinely entangle demographic with identity structure, motivating research into
objectives that disentangle them.

\textbf{Embedding geometry, not leakage magnitude, shapes suppressibility.}
The central finding of this work is that initial leakage magnitude is a poor predictor
of how much that leakage can be removed post-hoc.
Voice embeddings carry the highest baseline demographic leakage in the study
(86.1\% for ECAPA-TDNN, 94.8\% for WavLM), yet are the most suppressible modality:
AED reduces ECAPA-TDNN to 52.1\% and WavLM to 57.0\% with near-zero authentication
cost, recovering 94\% of the theoretical suppression ceiling.
Keystroke and touchstroke embeddings, by contrast, show only moderate baseline leakage
(62 - 69\%) yet resist suppression across all four methods; IVE degrades EER to
35-45\% for marginal privacy gains in several settings.
The PIC heatmap (\cref{fig:heatmap}) captures this asymmetry starkly: voice columns
are the deepest blue in the study while most keystroke columns cluster near zero or turn
red.

The explanation lies in the geometric structure of each embedding space rather than in
its leakage level. When demographic information occupies a separable subspace-as in voice
embeddings and in several gaze and gait convolutional models-post-hoc suppression can
identify and remove that subspace while leaving identity-discriminative directions
intact. When demographic and identity structure are geometrically entangled-as in
keystroke/touchstroke embeddings where timing and pressure signals carry both user
identity and body characteristics-any operation that reduces demographic separability
simultaneously degrades authentication utility.
This geometry-first view has a direct practical implication: measuring an
already-deployed model's demographic leakage level is necessary but insufficient.
The geometric organization of that leakage must also be characterized, for example
via linear separability probes or principal-direction analysis, before a post-hoc
suppression strategy can be responsibly recommended.

\textbf{Pareto frontier shape encodes the privacy--utility exchange rate.}  Ideal suppression in the Pareto curve would be shifting a dashed line vertically down without moving horizontally; however due to the privacy-utility trade off, there is usually a shift in horizontal axis. ~\Cref{fig:pvi_all_modalities} shows that gaze and gait often gain
privacy at near-zero EER cost, while keystroke/touchstroke reaches EER $\approx$46\%
for modest privacy gains.

% Chart 3 legend colors
\definecolor{clrGaze}{HTML}{2196F3}
\definecolor{clrGait}{HTML}{4CAF50}
\definecolor{clrVoice}{HTML}{F44336}
\definecolor{clrKeys}{HTML}{8931EF}
\definecolor{clrMeth}{HTML}{555555}
\definecolor{clrRand}{HTML}{E53935}

  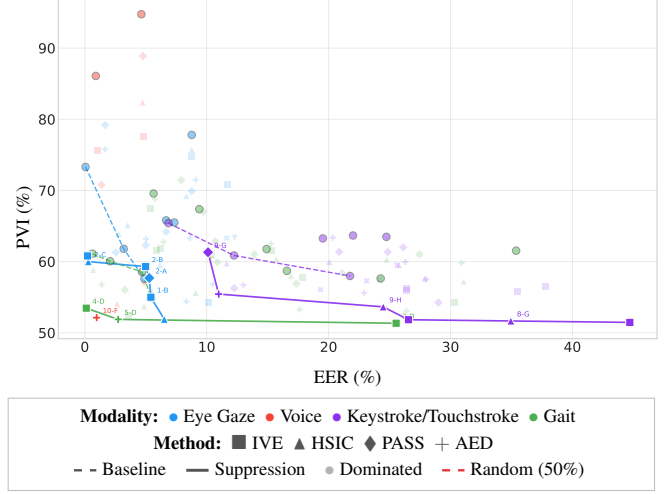
\begin{figure}[!t]
  \centering
  %-- plot with LaTeX axis labels
  \begin{tikzpicture}
    \node[inner sep=0pt] (img)
      {\includegraphics[width=\linewidth]{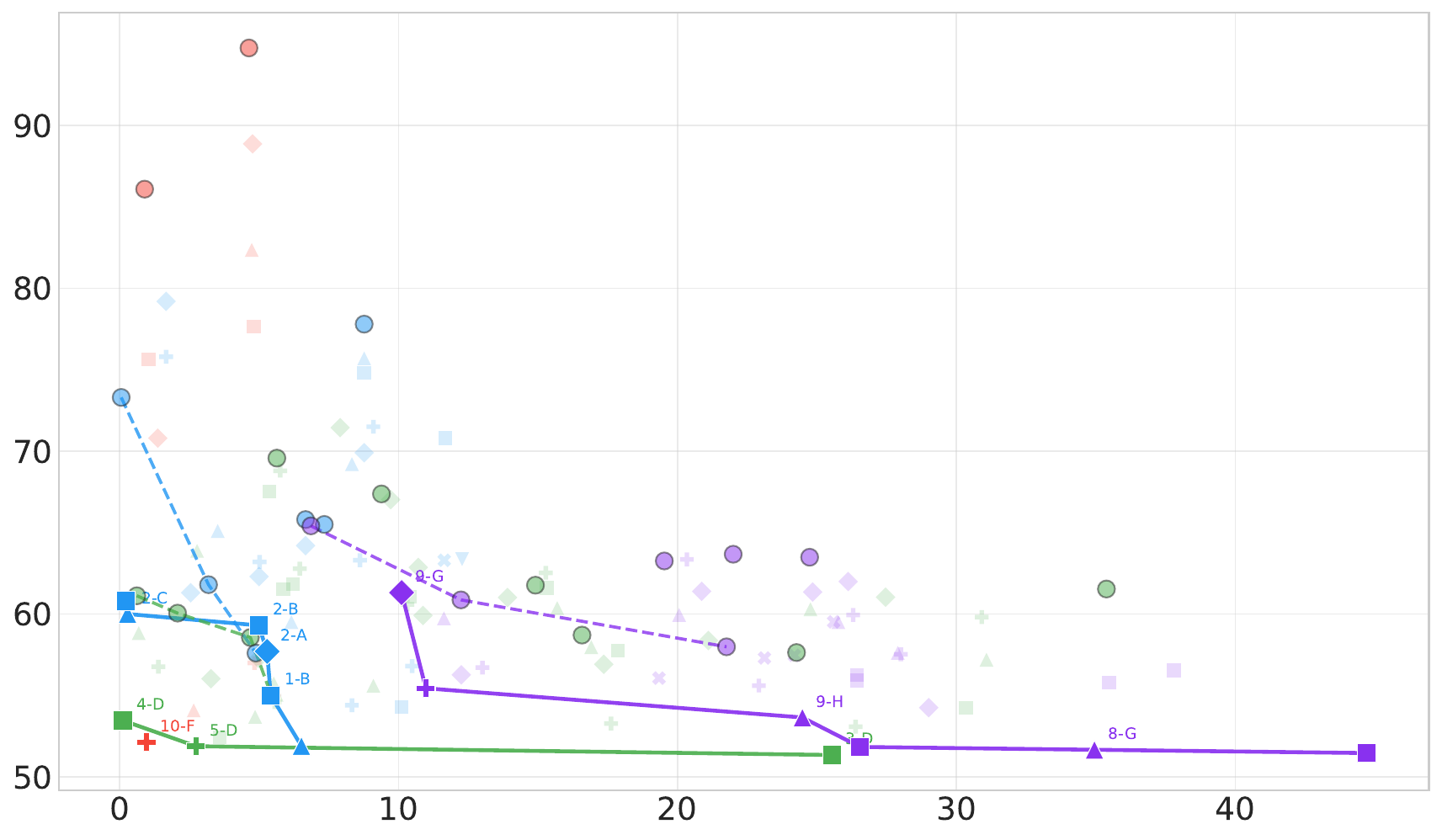}};
    \node[below=0pt of img.south, font=\scriptsize] {EER (\%)};
    \node[rotate=90, left=1mm of img.west, font=\scriptsize] {PVI (\%)};
  \end{tikzpicture}   

\setlength{\fboxsep}{4pt}
\setlength{\fboxrule}{0.4pt}

\noindent
\fcolorbox{gray!50}{white}{%
\parbox{0.98\columnwidth}{%
\centering\scriptsize

\textbf{Modality:}\enspace
\textcolor{clrGaze}{$\bullet$}~Eye Gaze\enspace
\textcolor{clrVoice}{$\bullet$}~Voice\enspace
\textcolor{clrKeys}{$\bullet$}~Keystroke/Touchstroke\enspace
\textcolor{clrGait}{$\bullet$}~Gait

\vspace{2pt}

\textbf{Method:}\enspace
\textcolor{clrMeth}{$\blacksquare$}~IVE\enspace
\textcolor{clrMeth}{$\blacktriangle$}~HSIC\enspace
\textcolor{clrMeth}{$\blacklozenge$}~PASS\enspace
\textcolor{clrMeth}{$\mathbf{+}$}~AED\enspace

\vspace{2pt}

\tikz[baseline=-0.6ex]{\draw[clrMeth,dashed,line width=0.9pt](0,0)--+(1.2em,0);}~Baseline\quad
\tikz[baseline=-0.6ex]{\draw[clrMeth,line width=1.1pt](0,0)--+(1.2em,0);}~Suppression\quad
{\color{gray!60}$\bullet$}~Dominated\quad
\tikz[baseline=-0.6ex]
{\draw[clrRand,dashed,line width=0.9pt](0,0)--+(1.2em,0);}~Random (50\%)

}
}
  \vspace{1pt}
  \caption{Privacy Vulnerability Index (PVI) vs.\ authentication error (EER)
  pareto curves.}
  \label{fig:pvi_all_modalities}
  \end{figure}

%--------------------------------------------------------------------
\section{Conclusion}
%--------------------------------------------------------------------
We present the first systematic multi-modal audit of demographic privacy leakage and
post-hoc suppression across behavioral biometric embeddings spanning eye gaze, voice,
keystroke/touchstroke, and gait. Leakage appears in every tested model: metric-learning
objectives and stronger authentication training tend to encode more demographic structure
into the embedding space. Post-hoc suppression methods originally developed for face
and speaker verification transfer successfully across behavioral biometric modalities,
but their effectiveness varies substantially across settings.

The central finding is that \textbf{demographic suppressibility is better explained by embedding geometry than by initial leakage magnitude alone}. Voice embeddings, despite
the highest baseline leakage (up to 94.8\%), were the most suppressible, while
keystroke and touchstroke embeddings with moderate leakage (62-69\%) resisted
suppression even at high authentication cost. Thus, the practical question shifts from \emph{``how much does this model leak?''}
to \emph{``how is that leakage geometrically organized?''} Audits should therefore
measure geometric separability, not leakage level alone.

\clearpage
\balance
% References (manual, for submission)


{\small
\begin{thebibliography}{99}

\bibitem{eberz2015}
S.~Eberz, K.~Rasmussen, V.~Lenders, and I.~Martinovic,
``Preventing lunchtime attacks: Fighting insider threats with eye movement
biometrics,''
in \textit{Proc. NDSS}, Internet Society, 2015.

\bibitem{kroger2020}
J.~L. Kr{\"o}ger, O.~H.-M. Lutz, and F.~M{\"u}ller,
``What does your gaze reveal about you? On the privacy implications of eye
tracking,''
in \textit{IFIP Summer School on Privacy and Identity Management},
Springer, 2020, pp.~226--241.

\bibitem{liu2019}
A.~Liu, L.~Xia, A.~Duchowski, R.~Bailey, K.~Holmqvist, and E.~Jain,
``Differential privacy for eye-tracking data,''
in \textit{Proc. ACM ETRA}, 2019, pp.~1--10.

\bibitem{makowski2020}
S.~Makowski, L.~A. J{\"a}ger, P.~Prasse, and T.~Scheffer,
``Biometric identification and presentation-attack detection using micro-
and macro-movements of the eyes,''
in \textit{Proc. IJCB}, IEEE, 2020, pp.~1--10.

\bibitem{abdrabou2024}
Y.~Abdrabou, M.~Hassib, S.~Hu, K.~Pfeuffer, M.~Khamis, A.~Bulling, and
F.~Alt,
``Eyeseeidentity: Exploring natural gaze behavior for implicit user
identification during photo viewing,'' 2024.

\bibitem{zhang2018}
Y.~Zhang, W.~Hu, W.~Xu, C.~T. Chou, and J.~Hu,
``Continuous authentication using eye movement response of implicit visual
stimuli,''
\textit{Proc. ACM IMWUT}, vol.~1, no.~4, pp.~1--22, 2018.

\bibitem{makowski2021}
S.~Makowski, P.~Prasse, D.~R. Reich, D.~Krakowczyk, L.~A. J{\"a}ger, and
T.~Scheffer,
``DeepEyedentificationLive: Oculomotoric biometric identification and
presentation-attack detection using deep neural networks,''
\textit{IEEE Trans. TBIOM}, vol.~3, no.~4, pp.~506--518, 2021.

\bibitem{lohr2020}
D.~Lohr, H.~Griffith, S.~Aziz, and O.~Komogortsev,
``A metric learning approach to eye movement biometrics,''
in \textit{Proc. IJCB}, IEEE, 2020, pp.~1--7.

\bibitem{lohr2022a}
D.~Lohr, H.~Griffith, and O.~V. Komogortsev,
``Eye know you: Metric learning for end-to-end biometric authentication
using eye movements from a longitudinal dataset,''
\textit{IEEE Trans. TBIOM}, vol.~4, no.~2, pp.~276--288, 2022.

\bibitem{lohr2022b}
D.~Lohr and O.~V. Komogortsev,
``Eye know you too: Toward viable end-to-end eye movement biometrics for
user authentication,''
\textit{IEEE Trans. IFS}, vol.~17, pp.~3151--3164, 2022.

\bibitem{raju2024}
M.~H. Raju, L.~Friedman, D.~J. Lohr, and O.~V. Komogortsev,
``Temporal persistence and intercorrelation of embeddings learned by an
end-to-end deep learning eye movement-driven biometrics pipeline,''
\textit{arXiv:2402.16399}, 2024.

\bibitem{yin2022}
J.~Yin, J.~Sun, J.~Li, and K.~Liu,
``An effective gaze-based authentication method with the spatiotemporal
feature of eye movement,''
\textit{Sensors}, vol.~22, no.~8, p.~3002, 2022.

\bibitem{eberz2016}
S.~Eberz, K.~B. Rasmussen, V.~Lenders, and I.~Martinovic,
``Looks like eve: Exposing insider threats using eye movement biometrics,''
\textit{ACM Trans. Privacy Security}, vol.~19, no.~1, pp.~1--31, 2016.

\bibitem{eberz2017}
S.~Eberz, K.~B. Rasmussen, V.~Lenders, and I.~Martinovic,
``Evaluating behavioral biometrics for continuous authentication: Challenges
and metrics,''
in \textit{Proc. AsiaCCS}, 2017, pp.~386--399.

\bibitem{bozkir2023}
E.~Bozkir, S.~{\"O}zdel, M.~Wang, B.~David-John, H.~Gao, K.~Butler,
E.~Jain, and E.~Kasneci,
``Eye-tracked virtual reality: A comprehensive survey on methods and privacy
challenges,''
\textit{arXiv:2305.14080}, 2023.

\bibitem{grandhi2025}
S.~G. Grandhi and S.~Samet,
``Evaluating the long-term viability of eye-tracking for continuous
authentication in virtual reality,''
\textit{arXiv:2502.20359}, 2025.

\bibitem{steil2019}
J.~Steil, I.~Hagestedt, M.~X. Huang, and A.~Bulling,
``Privacy-aware eye tracking using differential privacy,''
in \textit{Proc. ACM ETRA}, 2019, pp.~1--9.

\bibitem{jia2018}
S.~Jia, A.~Seccia, P.~Antonenko, R.~Lamb, A.~Keil, M.~Schneps, and
M.~Pomplun,
``Biometric recognition through eye movements using a recurrent neural
network,''
in \textit{Proc. IEEE ICBK}, 2018, pp.~57--64.

\bibitem{behaveformer2023}
D.~Senarath, S.~Tharinda, M.~Vishvajith, S.~Rasnayaka,
S.~Wickramanayake, and D.~Meedeniya,
``BehaveFormer: A framework with spatio-temporal dual attention transformers
for IMU-enhanced keystroke dynamics,''
in \textit{Proc. IJCB}, IEEE, 2023, pp.~1--9.

\bibitem{senarath2023}
D.~Senarath \textit{et al.},
``Re-evaluating keystroke dynamics for continuous authentication,''
in \textit{Proc. ICARC}, IEEE, 2023, pp.~202--207.

\bibitem{dwgrl2023}
C.~Li \textit{et al.},
``Dynamic weighted gradient reversal network for visible-infrared person
re-identification,''
\textit{ACM Trans. Multimedia Comput. Commun. Appl.}, vol.~20, no.~1,
pp.~1--23, 2023.

\bibitem{dhar2021}
P.~Dhar, J.~Gleason, A.~Roy, C.~Castillo, and R.~Chellappa,
``PASS: Protected attribute suppression system for mitigating bias in face
recognition,''
in \textit{Proc. ICCV}, 2021.

\bibitem{terhorst2019}
P.~Terh{\"o}rst, N.~Damer, F.~Kirchbuchner, and A.~Kuijper,
``Suppressing gender and age in face templates using incremental variable
elimination,''
in \textit{Proc. ICB}, IEEE, 2019.

\bibitem{noe2020}
P.-G. No{\'e}, M.~Todisco, A.~Laperriere, J.~Patino,
N.~Evans, and T.~Kinnunen,
``Adversarial disentanglement of speaker representation for
attribute-driven privacy preservation,''
\textit{arXiv:2012.04454}, 2020.

\bibitem{bortolato2020}
B.~Bortolato, M.~Ivanovska, P.~Rot, J.~Krizaj, P.~Terh{\"o}rst,
N.~Damer, P.~Peer, and V.~Struc,
``Learning privacy-enhancing face representations through feature
disentanglement,''
in \textit{Proc. IEEE FG}, 2020, pp.~495--502.

\bibitem{gonzalez2025}
N.~Gonzalez \textit{et al.},
``Type2Branch: Keystroke biometrics based on a dual-branch architecture
with attention mechanisms and set2set loss,''
\textit{IEEE Trans. IFS}, 2025.

\bibitem{keyrecs2023}
T.~Dias \textit{et al.},
``KeyRecs: A keystroke dynamics and typing pattern recognition dataset,''
\textit{Data in Brief}, vol.~50, 2023.

\bibitem{ikdd2024}
I.~Tsimperidis \textit{et al.},
``IKDD: A keystroke dynamics dataset for user classification,''
\textit{Information}, vol.~15, no.~9, p.~511, 2024.

\bibitem{feta2022}
M.~Georgiev \textit{et al.},
``FETA: Fair evaluation of touch-based authentication,''
\textit{arXiv:2201.10606}, 2022.

\bibitem{nguyen2024}
K.-N. Nguyen \textit{et al.},
``Spatio-temporal dual-attention transformer for time-series behavioral
biometrics,''
\textit{IEEE Trans. TBIOM}, vol.~6, no.~4, pp.~591--601, 2024.

\bibitem{taha2023}
B.~Taha, S.~N.~A. Seha, D.~Y. Hwang, and D.~Hatzinakos,
``EyeDrive: A deep learning model for continuous driver authentication,''
\textit{IEEE J. STSP}, vol.~17, no.~3, pp.~637--647, 2023.

\bibitem{qin2025}
H.~Qin \textit{et al.},
``EMMixFormer: Mix transformer for eye movement recognition,''
\textit{IEEE Trans. Instrum. Meas.}, 2025.

\bibitem{desplanques2020}
B.~Desplanques, J.~Thienpondt, and K.~Demuynck,
``ECAPA-TDNN: Emphasized channel attention, propagation and aggregation
in TDNN based speaker verification,''
in \textit{Proc. Interspeech}, 2020, pp.~3830--3834.

\bibitem{chen2022wavlm}
S.~Chen \textit{et al.},
``WavLM: Large-scale self-supervised pre-training for full stack speech
processing,''
\textit{IEEE J. Sel. Topics Signal Process.}, vol.~16, no.~6,
pp.~1505--1518, 2022.

\bibitem{nagrani2017}
A.~Nagrani, J.~S. Chung, and A.~Zisserman,
``VoxCeleb: A large-scale speaker identification dataset,''
in \textit{Proc. Interspeech}, 2017, pp.~2616--2620.

\bibitem{ngo2015}
T.~T.~Ngo, Y.~Makihara, H.~Nagahara, Y.~Mukaigawa, and Y.~Yagi,
``Similar gait action recognition using an inertial sensor,''
\textit{Pattern Recognit.}, vol.~48, no.~4, pp.~1289--1301, 2015.

\bibitem{zou2020}
Q.~Zou, Y.~Wang, Q.~Wang, Y.~Zhao, and Q.~Li,
``Deep learning-based gait recognition using smartphones in the wild,''
\textit{IEEE Trans. Inf. Forensics Security}, vol.~15, pp.~3197--3212, 2020.

\bibitem{antal2020}
M.~Antal and R.~Gy\"or\"odi,
``FeatureGait: A new feature extraction approach for gait-based person identification using convolutional neural networks,''
in \textit{Proc. Int. Conf. Artificial Neural Networks (ICANN)}, 2020, pp.~595--606.

\bibitem{delgadosantos2023}
P.~Delgado-Santos, R.~Tolosana, R.~Guest, F.~Deravi, and R.~Vera-Rodriguez,
``Exploring transformers for behavioural biometrics: A case study in gait recognition,''
\textit{Pattern Recognit.}, vol.~143, Art.~no.~109798, 2023.

\bibitem{ngo2014}
T.~T.~Ngo, Y.~Makihara, H.~Nagahara, Y.~Mukaigawa, and Y.~Yagi,
``The largest inertial sensor-based gait database and performance evaluation of gait-based personal authentication,''
\textit{Pattern Recognit.}, vol.~47, no.~1, pp.~222--231, 2014.

\bibitem{nusFeatureGait020}
S.~M.~C.~Terence,
``NUS IMU Gait Dataset,''
ScholarBank@NUS Repository, Dataset, Sep.~2020. [Online]. Available: https://doi.org/10.25540/3TCB-6594

\bibitem{rasnayaka2020}
S.~Rasnayaka and T.~Sim,
``Your tattletale gait: Privacy invasiveness of IMU gait data,''
in \textit{Proc. IEEE Int. Joint Conf. Biometrics (IJCB)}, Houston, TX, USA, 2020, pp.~1--10.

\bibitem{gazebasevr2022}
D.~Lohr, S.~Aziz, L.~Friedman, and O.~Komogortsev,
``GazeBaseVR Data Repository,''
figshare, Dataset, 2022. [Online]. Available: https://doi.org/10.6084/m9.figshare.21308391.v2

\bibitem{gazebase2020}
H.~Griffith, D.~Lohr, and O.~V.~Komogortsev,
``GazeBase Data Repository,''
figshare, Dataset, 2020. [Online]. Available: https://doi.org/10.6084/m9.figshare.12912257.v3

\bibitem{hanisch2025}
S.~Hanisch, P.~Arias-Cabarcos, J.~Parra-Arnau, and T.~Strufe,
``Anonymization techniques for behavioral biometric data: A survey,''
\textit{ACM Comput. Surv.}, vol.~57, no.~11, Art.~272, Jun. 2025.
\doi{10.1145/3729418}

\bibitem{melzi2024}
P.~Melzi, C.~Rathgeb, R.~Tolosana, R.~Vera-Rodr{\'i}guez, and C.~Busch,
``An overview of privacy-enhancing technologies in biometric recognition,''
\textit{ACM Comput. Surv.}, vol.~56, no.~12, Art.~310, Oct. 2024.
\doi{10.1145/3664596}

\end{thebibliography}
}
\end{document}